\documentstyle[12pt]{article}
\newcommand{\dfrac}[2]{\displaystyle{\frac{#1}{#2}}}
\newcommand{\dsum}[2]{\displaystyle{\sum_{#1}^{#2}}}
\newcommand{\dprod}[2]{\displaystyle{\prod_{#1}^{#2}}}
\newcommand{\dint}[2]{\displaystyle{\int_{#1}^{#2}}}
\newcommand{\overl}[1]{\overline{#1}}

\title{Bethe Ansatz Equations \\ for \\ the
       Broken $Z_{N}$-Symmetric Model}
\author{Yuji Yamada \\
        {\it Research Institute for Mathematical Sciences}\\
        {\it Kyoto University} \\
        {\it Sakyo-ku, Kyoto, 606-01, Japan}\\
        ${\bf e}$-mail:\  yamada@kurims.kyoto-u.ac.jp}
\date{}
\begin {document}
\maketitle

\begin{abstract}
We obtain the Bethe Ansatz equations for the broken
${\bf Z}_{N}$-symmetric model by
constructing a functional relation of the transfer matrix
of $L$-operators.
This model is an elliptic off-critical extension of the
Fateev-Zamolodchikov model.
We calculate the free energy of this model
on the basis of the string hypothesis.

\vspace*{3mm}

\end{abstract}

\section{Introduction} \label{sec:intro}

In the two dimensional  solvable lattice models
with Ising-like edge interaction,
the star-triangle relation
\begin{eqnarray}
 \lefteqn{\rho \,W(a,b|v,w)\,\overline{W}(a,c|u,w)\,W(b,c|u,v)} \nonumber\\
 &=& \sum_{d}
  \overline{W}(a,d|u,v)\,W(d,b|u,w)\,\overline{W}(d,c|v,w)
 \label{eq:STRintro}
\end{eqnarray}
\begin{displaymath}
 \rho=\rho(u,v,w)\quad\mbox{independent of $a$, $b$ and $c$}
\end{displaymath}
plays a central role.
In (\ref{eq:STRintro}), the summation on $d$ is taken over all local states.
These are the relations among the two Boltzmann weights $W(a,b|u,v)$
and $\overline{W}(a,b|u,v)$.
They live on the edges in two different directions
of the two dimensional planar lattice.
The local state variables $a$ and $b$ live on the sites.
We denote the spectral parameters by $u$ and $v$.
$$(Fig.\  1\quad\mbox{and}\quad Fig.\ 2)$$

Since Fateev and Zamolodchikov \cite{FZ82}
obtained an $N$-state generalization of the critical Ising model
as a solution
of the star-triangle relation (STR),
there have been known two different off-critical extensions
of this model. One is the chiral Potts model \cite{AMcPTY87}
and the other is the broken ${\bf Z}_{N}$-symmetric model.
Both are Ising-type edge interaction models.
The STR for the chiral Potts model
was proved in \cite{BPAuY88}\cite{AuYP89}.
Though this model has been still under investigation in
\cite{BBP89}\nocite{Baxter88-2}\nocite{AMcP89-3}\nocite{Baxter90}
\nocite{ADMc91}--\cite{Albertini92},
the lack of a difference-variable parameterization in this model
causes difficulties in analysis.
Kashiwara and Miwa \cite{KM86} proposed  the broken
${\bf Z}_{N}$-symmetric model,
and Hasegawa and Yamada \cite{HY90} proved the STR for this model.
Unfortunately the proof in \cite{KM86} was wrong
because of the incorrectness of the ``ICU lemma'' in their paper.

In this Paper, we study the eigenvalues of the transfer matrix
$\Phi(u,v,w)$ of the broken ${\bf Z}_{N}$-symmetric model,
\begin{equation}
 \Phi(u,v,w)^{b_{0}b_{1}\cdots b_{M-1}}_{a_{0}a_{1}\cdots a_{M-1}}
 =
 \prod_{j=0}^{M-1}\overline{W}(b_{j},a_{j}|v-w)\,W(a_{j},b_{j+1}|u-w),
 \label{eq:defPHI}
\end{equation}
and calculate the free energy of this model.
$$(Fig.\ 3)$$
The local state variables take their values
in ${\bf Z}/N{\bf Z}$.
Throughout the Paper, we deal with the case of $N$ odd, $N=2n+1$.
The ${\bf Z}_{2}$-symmetry of the Boltzmann weights
\begin{equation}
 W(a,b|u)=W(N-a,N-b|u),\quad
 \overl{W}(a,b|u)=\overl{W}(N-a,N-b|u),
\end{equation}
ensures that the eigenvalue $r=\pm 1$ of the spin reversal operator
${\cal R}$ is a good quantum number,
where ${\cal R}$ $\in $ ${\rm End}\,(({\bf C}^{N})^{\otimes M})$ is defined by
\begin{equation}
 {\cal R}=\overbrace{R \otimes R \otimes \cdots \otimes R}
          ^{M\ times},
 \quad
 R\,v_{j}^{(N)}=v_{N-j}^{(N)},
 \label{eq:defR}
\end{equation}
which satisfies ${\cal R}^{2}=1$.
The vectors $v^{(N)}_{j}$ ($j\in {\bf Z}/N{\bf Z}$)
constitute an orthonormal basis in ${\bf C}^{N}$.
In the homogeneous case $u=v$,
we show first that any eigenvalue $\varphi(u)$ of $\Phi(u)=\Phi(u,u,0)$
can be written as
\begin{eqnarray}
 &
 \varphi(u)=\Biggl(
            \dfrac{p(0)p(\lambda)}{p(u)p(\lambda-u)}
            \Biggr)^{M}\,
            \dprod{j=1}{2nM}
            \dfrac{\theta_{1}(u-u_{j}|\tau/2)}
                  {\theta_{1}(u_{j}|\tau/2)},
 & \label{eq:introvarphi}\\
 &
 p(u) = \dprod{j=1}{n}\theta_{2}(u-(2j-1)\eta|\tau/2),
 \quad \eta=\dfrac{n}{N},\quad\lambda=\dfrac{1}{2}-\eta.
 & \label{eq:introp}
\end{eqnarray}
See Appendix A for the notation of the theta functions.
The zeros $\{u_{1},\cdots,u_{2nM}\}$ of $\varphi(u)$  are described
as follows
\begin{eqnarray}
 &
 \Biggl(
 \dfrac{\theta_{1}(v_{k}+\lambda/2|\tau/2)}
       {\theta_{1}(v_{k}-\lambda/2|\tau/2)}
 \Biggr)^{2M}
 =(-1)^{M+1}
 \displaystyle{\prod_{j=1}^{2nM}}
 \dfrac{\theta_{1}(v_{k}-v_{j}+\eta|\tau/2)}
       {\theta_{1}(v_{k}-v_{j}-\eta|\tau/2)},
 &\label{eq:introBAE} \\
 &
 v_{k}=u_{k}-\dfrac{\lambda}{2}
 \quad \mbox{for}\quad k=1,\cdots,2nM,
 & \label{eq:introuv}\\
 &
 \displaystyle{\sum_{j=1}^{2nM}}v_{j} \equiv
 \dfrac{1-r}{4}\  \bmod\bigl({\bf Z} \oplus \dfrac{\tau}{2}{\bf Z}).
 \label{eq:sumcond}
 &
\end{eqnarray}
We call the equation (\ref{eq:introBAE}) the Bethe Ansatz equation.
The condition (\ref{eq:sumcond}) follows from the
double periodicity of $\varphi(u)$ discussed in Section 4.
We obtain the Bethe Ansatz equations above through a functional
relation (\ref{eq:introFR}) for the transfer matrix of $L$-operators.
These $L$-operators
$L(u)$ $\in$ $\mbox{End}({\bf C}^{N} \otimes{\bf C}^{2})$
were originally constructed by Sklyanin
\cite{Sklyanin82-1}\cite{Sklyanin82-2} as a solution to
the relation,
\begin{eqnarray}
 \lefteqn{
 L^{01}(u-v)\,L^{02}(u-w)\,R^{12}_{8V}(v-w)} \nonumber \\
 &=&
 R^{12}_{8V}(v-w)\,L^{02}(u-w)\,L^{01}(u-v) \hspace{5mm}
 \mbox{on} \hspace{5mm}
 {\bf C}^{N} \otimes {\bf C}^{2} \otimes {\bf C}^{2},
 \label{eq:LLR}
\end{eqnarray}
$$(Fig.\ 4)$$
where the upper indices $0$, $1$ and $2$ mean that
$L^{ij}(u)$ acts only on the $i$-th and $j$-th
components of ${\bf C}^{N} \otimes {\bf C}^{2} \otimes {\bf C}^{2}$
and as identity on the other component.
We denote the $R$-matrix of the eight-vertex model by $R_{8V}(u)$
\cite{Baxter71}\cite{Baxter72}.
We consider the transfer matrix ${\cal L}(u)$ of these $L$-operators,
\begin{eqnarray}
 &
 {\cal L}(u)
  = tr_{{\bf C}^{2}}
  (L^{0M}(u)L^{1M}(u)\cdots L^{M-1\,M}(u)),
 & \\
 &
 L^{0M}(u)L^{1M}(u)\cdots L^{M-1\,M}(u) \in
 {\rm End}(\overbrace{{\bf C}^{N}\otimes\cdots\otimes{\bf C}^{N}}
      ^{\scriptscriptstyle{M \  times}}\otimes{\bf C}^{2}).
 &
\end{eqnarray}
We derive the functional relation
\begin{eqnarray}
 &
 \kern-2em {\cal L}(\lambda-u-1/4)\,\Phi(u)
 =
 C(u)^{M}
 \Biggl(f(u)^{M}\Phi(u-\eta)+\Bigl(-f(\lambda-u)\Bigr)^{M}\Phi(u+\eta)\Biggr),
 & \label{eq:introFR}\\
 &
 f(u)=\theta_{1}(2u|\tau)\,\dfrac{\theta_{1}(u+\eta|\tau/2)}
                                 {\theta_{2}(u|\tau/2)},
 & \label{eq:introf}\\
 &
 C(u)={[}\theta_{2}\theta_{3}\theta_{4}{]}(0|\tau)
      \dprod{j=1}{n}
      \dfrac{\theta_{1}(u-2(j-1)\eta|\tau/2)\,\theta_{2}(u+(2j-1)\eta|\tau/2)}
            {\theta_{1}(u+2j\eta|\tau/2)\,\theta_{2}(u-(2j-1)\eta|\tau/2)},
 & \label{eq:introC}
\end{eqnarray}
by the method which Baxter employed to solve the eight-vertex
model \cite{Baxter71}\cite{Baxter72}\cite{Bible82}.
The functional relations corresponding in the chiral Potts
model and in the RSOS model associated with the
eight-vertex were obtained in \cite{BS90} and \cite{BR89},
respectively.

We calculate the free energy
of the broken ${\bf Z}_{N}$-symmetric model
under the hypothesis that the solution of the Bethe Ansatz
equations corresponding to the ground state
consists of ``strings of length $N-1$''.
We show
that, in the infinite lattice limit, the centers of these strings are
distributed on the imaginary axis with the density $\rho(w)$,
\begin{equation}
 \rho(w)
 =
 2N\,{[}\theta_{2}\theta_{3}{]}\Bigl(0|N\tau\Bigr)\,
 \dfrac{\theta_{3}\Bigl(2\sqrt{-1}Nw|N\tau\Bigr)}
       {\theta_{2}\Bigl(2\sqrt{-1}Nw|N\tau\Bigr)},
 \quad
 -\dfrac{\kappa}{4}\leq w < \dfrac{\kappa}{4},
\end{equation}
where $\tau=\sqrt{-1}\kappa$,
and the free energy per site is
\begin{equation}
 F(u)
 =
 -\sum_{l=1}^{\infty}
 \dfrac{\sinh\Bigl(\dfrac{2\pi l}{\kappa}u\Bigr)\,
        \sinh\Bigl(\dfrac{2\pi l}{\kappa}(\dfrac{1}{2N}-u)\Bigr)\,
        \sinh\Bigl(\dfrac{2\pi l}{N\kappa}n\Bigr)}
       {l\,\cosh\Bigl(\dfrac{\pi l}{\kappa}\Bigr)\,
           \cosh^{2}\Bigl(\dfrac{\pi l}{N\kappa}\Bigr)}.
\end{equation}
The last expression agrees with the result of Jimbo, Miwa and Okado
\cite{JMO86} obtained by the use of the inversion-trick,
and in the trigonometric limit
of $\kappa \rightarrow \infty$ it recovers the result of
Fateev and Zamolodchikov \cite{FZ82} and Albertini \cite{Albertini92}.

The organization of this Paper is as follows.
In Section 2, we review necessary facts about the $R$-matrix of the
eight-vertex model, Sklyanin's $L$-operator and the broken
${\bf Z}_{N}$-symmetric model.
We derive the functional relation (\ref{eq:introFR}) in Section 3.
After showing commutation relations
among $\Phi(u)$, ${\cal L}(v)$ and ${\cal R}$,
we obtain the Bethe Ansatz equations in Section 4.
We calculate the free energy of the model
under the string hypothesis in Section 5.
Finally, in Section 6, we conclude with a brief discussion.
We fix the notation and list the formulae for theta functions
in Appendix \ref{ap:theta}.
Miscellaneous properties of the Boltzmann weights
are summarized in Appendix B.
We devote Appendix C to the proof of the commutativity between $\Phi$
and ${\cal L}$.

\section{Review of the Broken ${\bf Z}_{N}$ Symmetric Model}
\setcounter{equation}{0}
\label{sec:review}

We fix the notation for matrices.
We denote the vector
$\mbox{}^{t}(0,\cdots,\stackrel{j}{\check{1}},\cdots,0)$
in ${\bf C}^{m}$ by $v_{j}^{(m)}$, $j=0,1,\cdots,m-1$,
and the matrix elements of
$A \in {\rm End}({\bf C}^{m_{1}}\otimes{\bf C}^{m_{2}}\otimes
 \cdots\otimes{\bf C}^{m_{l}})$ by
\begin{displaymath}
 A\,v_{i_{1}}\otimes v_{i_{2}}\otimes\cdots\otimes v_{i_{l}}
 =
 \dsum{j_{1}=0}{m_{1}-1}\dsum{j_{2}=0}{m_{2}-1}
 \cdots\dsum{j_{l}=0}{m_{l}-1}
 v_{j_{1}}\otimes v_{j_{2}}\otimes\cdots\otimes v_{j_{l}}\,
 A^{j_{1}j_{2}\cdots j_{l}}_{i_{1}i_{2}\cdots i_{l}}.
\end{displaymath}
In \cite{Sklyanin82-1}\cite{Sklyanin82-2},
Sklyanin constructed the $L$-operators
$L(u)$ $\in$ $\mbox{End}({\bf C}^{N} \otimes{\bf C}^{2})$ for
the eight vertex model satisfying (\ref{eq:LLR}).
The $R$-matrix of the eight-vertex model
$R_{8V}(u)$ \cite{Baxter71}\cite{Baxter72} is
a solution of the Yang-Baxter equation,
\begin{eqnarray}
 \lefteqn{
 R^{01}_{8V}(u-v)\,R^{02}_{8V}(u-w)\,R^{12}_{8V}(v-w)} \nonumber \\
 &=&
 R^{12}_{8V}(v-w)\,R^{02}_{8V}(u-w)\,R^{01}_{8V}(u-v) \hspace{5mm}
 \mbox{on} \hspace{5mm}
 {\bf C}^{2} \otimes {\bf C}^{2} \otimes {\bf C}^{2}.
\end{eqnarray}
Its non-zero matrix elements are
\begin{eqnarray*}
 &
 R\mbox{}^{00}_{00}(u) = R\mbox{}^{11}_{11}(u) =
 {[}\theta_{2}\theta_{3}{]}(\eta){[}\theta_{2}\theta_{3}{]}(u)
 {[}\theta_{1}\theta_{4}{]}(u+\eta) ,
 & \\
 &
 R\mbox{}^{01}_{01}(u) = R\mbox{}^{10}_{10}(u) =
 {[}\theta_{2}\theta_{3}{]}(\eta){[}\theta_{1}\theta_{4}{]}(u)
 {[}\theta_{2}\theta_{3}{]}(u+\eta) ,
 & \\
 &
 R\mbox{}^{01}_{10}(u) = R\mbox{}^{10}_{01}(u) =
 {[}\theta_{1}\theta_{4}{]}(\eta){[}\theta_{2}\theta_{3}{]}(u)
 {[}\theta_{2}\theta_{3}{]}(u+\eta) ,
 & \\
 &
 R\mbox{}^{00}_{11}(u) = R\mbox{}^{11}_{00}(u) =
 {[}\theta_{1}\theta_{4}{]}(\eta){[}\theta_{1}\theta_{4}{]}(u)
 {[}\theta_{1}\theta_{4}{]}(u+\eta) .
 &
\end{eqnarray*}
$$(Fig.\ 5)$$
The other elements not specified above are all zero.
Here we denote $\theta_{1}(u)\theta_{4}(u)$
by ${[}\theta_{1}\theta_{4}{]}(u)$ for short.
We usually suppress the elliptic modulus $\tau$.
When $\eta=\dfrac{n}{N}$, $N=2n+1$, the $L$-operators $L(u)$
in (\ref{eq:LLR}) have a ``cyclic'' representation.
In this representation, $L(u)$ factorizes elementwise as
\begin{equation}
  L\mbox{}^{bj}_{ai}(u) = K\mbox{}^{\ b}_{ia}(u)\,K\mbox{}^{jb}_{\ a}(u),
 \label{eq:factorization}
\end{equation}
where
\begin{eqnarray}
 &
 K\mbox{}^{j\ a}_{\ a+\sigma}(u) = (-1)^{j(1+\sigma)/2}
 {[}\theta_{1+j}\theta_{4-j}{]}(u-\sigma a \eta +\frac{1}{4}),
 & \label{eq:Kexp1} \\
 &
  K\mbox{}^{\ b}_{ja}(u) = G_{a}^{-1}G_{b}^{-1}K\mbox{}^{ja}_{\ b}(u),
 \quad
  G_{a}=\Biggl(\dfrac{\theta_{4}(2a\eta)}{\theta_{4}(0)}\Biggr)^{1/2},
 &\label{eq:Kexp2}
\end{eqnarray}
for $j=0,1$, \  $a,b=0,1,\cdots,N-1$ and $\sigma=\pm1$.
$$(Fig.\ 6)$$
The factors $K\mbox{}_{ia}^{\ b}(u)$ and $K\mbox{}_{\ a}^{jb}(u)$
are zero unless $|a-b|=1$.
We can identify these $K(u)$'s as the intertwining vectors
appearing in the vertex-face correspondence
\cite{Baxter73-1}\nocite{Baxter73-2}\nocite{Baxter73-3}
\nocite{Pasquier88}\nocite{Hasegawa93}--\cite{Quano92}.
Even in the Fateev-Zamolodchikov model
these $K(u)$'s are different from the 3-spin object $V$'s
in \cite{AMcPTY87}\cite{BBP89} by definition.
Their $V$'s are defined by the Fourier transformed
images of the product of two Boltzmann weights.
These two objects, however, should have an intimate relationship,
because the transfer matrix ${\cal L}$ in the
Fateev-Zamolodchikov model is also constructed from $V$'s
\cite{BBP89}.

Under the ${\bf Z}_{2}$-transformation which sends
$a$ to $N-a$, they change as
\begin{equation}
 K\mbox{}^{j\,N-a}_{\ \,N-b}(u)=(-1)^{j(n-1)+1}K\mbox{}^{ja}_{\ b}(u),
 \quad
 K\mbox{}_{j\,N-b}^{\ \,N-a}(u)=(-1)^{j(n-1)+1}K\mbox{}_{jb}^{\ a}(u).
 \label{eq:Z2K}
\end{equation}
They satisfy the unitarity relations \cite{Hasegawa92}
\begin{eqnarray}
 \sum_{j=0}^{1}G_{a}
 K\mbox{}^{\ b}_{ja}(u+\lambda)\,K\mbox{}^{jb}_{\ c}(u)
 &=&\delta_{ac}
 {[}\theta_{2}\theta_{3}\theta_{4}{]}(0)\,G_{b}\,\theta_{2}(2u)
 \label{eq:unitarity1},\\
 \sum_{a=0}^{N-1}G_{a}
 K\mbox{}^{\ b}_{ia}(u+\lambda)\,K\mbox{}^{jb}_{\ a}(u)
 &=&\delta_{ij}
 {[}\theta_{2}\theta_{3}\theta_{4}{]}(0)\,G_{b}\,\theta_{2}(2u)
 \label{eq:unitarity2},\\
 \sum_{b=0}^{N-1}G_{b}
 K\mbox{}^{\ b}_{ia}(u)\,K\mbox{}^{jb}_{\ a}(u+\lambda)
 &=&\delta_{ij}
 {[}\theta_{2}\theta_{3}\theta_{4}{]}(0)\,G_{a}\,\theta_{2}(2u)
 \label{eq:unitarity3}.
\end{eqnarray}
$$(Fig.\ 7)$$
In \cite{HY90}, we determined the Boltzmann weights
$W$ and $\overline{W}$ of
the broken ${\bf Z}_{N}$- symmetric model
by the relations
\begin{eqnarray}
 \lefteqn{
 W(a,b|u,v) \sum_{j=0}^{1}K\mbox{}^{jc}_{\ a}(u-w)\,
                          K\mbox{}^{\ d}_{jb}(v-w)} \nonumber \\
 &=& \sum_{j=0}^{1}K\mbox{}^{jc}_{\ a}(v-w)\,
                   K\mbox{}^{\ d}_{jb}(v-w) \,W(c,d|u,v),
 \label{eq:defW}\\
 \lefteqn{
 \sum_{b=0}^{N-1}\overline{W}(a,b|u,v)\,
                 K\mbox{}^{\ c}_{ib}(u-w)\,
                 K\mbox{}^{jc}_{\ b}(v-w)} \nonumber \\
 &=& \sum_{b=0}^{N-1}K\mbox{}^{\ b}_{ia}(v-w)\,
                     K\mbox{}^{jb}_{\ a}(u-w)\,
                     \overline{W}(b,c|u,v).
 \label{eq:defWb}
\end{eqnarray}
$$(Fig.\ 8\quad\mbox{and}\quad Fig. 9)$$
{}From the above relations and (\ref{eq:Z2K}),
we have the ${\bf Z}_{2}$-symmetry
\begin{equation}
 W(a,b|u,v)=W(N-a,N-b|u,v),
 \quad
 \overl{W}(a,b|u,v)=\overl{W}(N-a,N-b|u,v).
\end{equation}
The equation (\ref{eq:defW}) implies that
$W(a,b|u,v)$$=$$W(a,b|u-v)$ and that
\begin{equation}
 \begin{array}{ccc}
 \displaystyle{\frac{W(a+1,b+1|u)}{W(a,b|u)}}
 &=& \displaystyle{
 \frac{\theta_{3}(u+(a+b+1)\eta)}{\theta_{3}(u-(a+b+1)\eta)}},\\
 \displaystyle{\frac{W(a+1,b-1|u)}{W(a,b|u)}}
 &=& \displaystyle{
 \frac{\theta_{2}(u+(a-b+1)\eta)}{\theta_{2}(u-(a-b+1)\eta)}}.
 \end{array}
 \label{eq:recW}
\end{equation}
The crossing symmetry
\begin{equation}
 \overline{W}(a,b|u)=G_{a}G_{b}W(a,b|\lambda-u)
 \label{eq:CroSym}
\end{equation}
holds in this model.
In the paper \cite{Hasegawa92}, Hasegawa proved
the crossing symmetry only from (\ref{eq:defW})
and the unitarity relations
(\ref{eq:unitarity2}) and (\ref{eq:unitarity3}).
We thus have
\begin{equation}
 \begin{array}{ccc}
 \displaystyle{\frac{\overline{W}(a+1,b+1|u)}{\overline{W}(a,b|u)}}
 &=& \displaystyle{
 \frac{G_{a+1}G_{b+1}}{G_{a}G_{b}}
 \frac{\theta_{4}(u-(a+b)\eta)}{\theta_{4}(u+(a+b+2)\eta)}},\\
 \displaystyle{\frac{\overline{W}(a+1,b-1|u)}{\overline{W}(a,b|u)}}
 &=& \displaystyle{
 \frac{G_{a+1}G_{b-1}}{G_{a}G_{b}}
 \frac{\theta_{1}(u-(a-b)\eta)}{\theta_{1}(u+(a-b+2)\eta)}},
 \end{array}
 \label{eq:recWb}
\end{equation}
without directly solving (\ref{eq:defWb}).
We can see from (\ref{eq:recW}) and (\ref{eq:recWb})
that the Boltzmann weights satisfy the reflection symmetry
\begin{displaymath}
 W(a,b|u) = W(b,a|u), \quad
 \overline{W}(a,b|u) = \overline{W}(b,a|u).
\end{displaymath}
Defining $T\mbox{}^{(+)}_{k}(\alpha|u)$
and $T\mbox{}^{(-)}_{k}(\alpha|u)$ by
\begin{equation}
 T\mbox{}^{(+)}_{k}(\alpha|u)=\prod_{j=1}^{\alpha}
             \displaystyle{\frac{\theta_{k}(u+(2j-1)\eta)}
                                {\theta_{k}(u-(2j-1)\eta)}}
 \quad\mbox{and}\quad
 T\mbox{}^{(-)}_{k}(\alpha|u)=T\mbox{}^{(+)}_{k}(\alpha|\lambda-u),
 \label{eq:T}
\end{equation}
respectively, the solutions to the recursion
relations (\ref{eq:recW}) and (\ref{eq:recWb})
under the normalization
$W(0,0|u)$ $=$ $\overline{W}(0,0|u)$ $=$ $1$ are
\begin{eqnarray*}
 W(2a,2b|u)
 &=&
 T^{(+)}_{2}(a-b|u)\,T^{(+)}_{3}(a+b|u), \\
 \overline{W}(2a,2b|u)
 &=&
 G_{2a}\,G_{2b}\,T^{(-)}_{2}(a-b|u)\,T^{(-)}_{3}(a+b|u).
\end{eqnarray*}
Here all local state variables are to be read modulo $N$.
See Appendix \ref{ap:Boltz} for details.
Hasegawa and Yamada in \cite{HY90}
established the star-triangle relation (STR)
in the broken ${\bf Z}_{N}$-symmetric model,
\begin{eqnarray}
 \lefteqn{\rho \,W(a,b|v-w)\,\overline{W}(a,c|u-w)\,W(b,c|u-v)} \nonumber \\
 &=& \sum_{d=0}^{N-1}
  \overline{W}(a,d|u-v)\,W(d,b|u-w)\,\overline{W}(d,c|v-w),
 \label{eq:STR}
\end{eqnarray}
where $\rho$ is a scalar function independent of $a$, $b$
and $c$.

\section{Functional Relation}\setcounter{equation}{0}
\label{sec:FuncRel}
In this section, we consider the transfer matrix of the $L$-operators
and construct a functional relation for it.
In the course of the calculation, we utilize the factorization property
of $L$ into $K$'s (\ref{eq:factorization}).
We define a 2-by-2 matrix $L(a,b,|u,v,w)$ by
\begin{displaymath}
L(a,b,|u,v,w) =
 \left( \begin{array}{cc}
        K\mbox{}^{\ b}_{0a}(u-w)\,K\mbox{}^{0b}_{\ a}(v-w)&
        K\mbox{}^{\ b}_{0a}(u-w)\,K\mbox{}^{1b}_{\ a}(v-w)\\
        K\mbox{}^{\ b}_{1a}(u-w)\,K\mbox{}^{0b}_{\ a}(v-w)&
        K\mbox{}^{\ b}_{1a}(u-w)\,K\mbox{}^{1b}_{\ a}(v-w)
 \end{array} \right).
\end{displaymath}
Then the transfer matrix ${\cal L}(u,v,w)$
of $L$-operators on the lattice of
width $M$ with the periodic boundary condition is
\begin{eqnarray}
 \lefteqn{{\cal L}(u,v,w)
          ^{b_{0}b_{1}\cdots b_{M-1}}_{a_{0}a_{1}\cdots a_{M-1}}}
 \nonumber \\
 &=& tr\bigg(L(a_{0},b_{0}|u,v,w)L(a_{1},b_{1}|u,v,w)\cdots
        L(a_{M-1},b_{M-1}|u,v,w)\bigg)
 \label{eq:tempdefL}  \\
 &=& \sum_{i_{0},\cdots,i_{M-1}}\prod_{j=0}^{M-1}
     K\mbox{}^{\ \ \ \ b_{j}}_{i_{j+1}a_{j}}(u-w)\,
     K\mbox{}^{i_{j} b_{j}}_{\ \ a_{j}}(v-w).
 \label{eq:defL}\nonumber
\end{eqnarray}
$$(Fig.\ 10)$$
The final goal of this section is to establish the functional relation
\begin{eqnarray}
 \lefteqn{{\cal L}(\lambda-u,\lambda-v,w+1/4)\,
           \Phi(u,v,w)} \nonumber \samepage \\
 &=& C(u,v,w)^{M}
 \left(\begin{array}{l}
   f(u,v,w)^{M}\Phi(u,v,w+\eta) \\
   \qquad\qquad +\Bigl(-f(\lambda-v,\lambda-u,-w)\Bigr)^{M}\Phi(u,v,w-\eta)
 \end{array}\right),
 \label{eq:matFR}
\end{eqnarray}
where we define $C(u,v,w)$ and $f(u,v,w)$ by
\begin{equation}
 C(u,v,w)={[}\theta_{2}\theta_{3}\theta_{4}{]}(0)\,
          {[}T^{(+)}_{2}T^{(+)}_{3}{]}(n|u-w)\,
          {[}T^{(-)}_{2}T^{(-)}_{3}{]}(n|v-w),
 \label{eq:C}
\end{equation}
\begin{equation}
 f(u,v,w)
 =
 \theta_{1}(2u-2w)\,
 \displaystyle{\frac{{[}\theta_{1}\theta_{4}{]}(v-w+\eta)}
                    {{[}\theta_{2}\theta_{3}{]}(u-w)}}.
 \label{eq:newf}
\end{equation}
This functional relation reduces to (\ref{eq:introFR}),
(\ref{eq:introf}) and (\ref{eq:introC}) in the homogeneous
case, $u=v$.
We achieve this goal by the method \`a la Baxter
\cite{Baxter71}\cite{Baxter72}\cite{Bible82},
which states the following:
Suppose that we can find ${\bf C}$-valued functions
$\phi_{j}^{(\nu)}(b|u,v,w)$ ($\nu=0,1,2,3$ and $b\in{\bf Z}/N{\bf Z}$)
and matrices $P_{j}$ $\in$ ${\rm End}\,({\bf C}^{2})$
for $j\in{\bf Z}/M{\bf Z}$, which satisfy
\begin{eqnarray}
 \lefteqn{
 P_{j}^{-1}\dsum{b=0}{N-1}\phi_{j}^{(0)}(b|u,v,w)\,
          L(a,b|u,v,w) \,P_{j+1}} \nonumber \\
 &=& \left( \begin{array}{cc}
               \phi_{j}^{(1)}(a|u,v,w) & \phi_{j}^{(3)}(a|u,v,w) \\
                         0             & \phi_{j}^{(2)}(a|u,v,w)
   \end{array}\right)
 \quad\mbox{for}\ a \in{\bf Z}/N{\bf Z}
 \ \mbox{and}\   j\in{\bf Z}/M{\bf Z}.
 \label{eq:phiP}
\end{eqnarray}
Then we have
\begin{eqnarray}
 &
 \dsum{b_{0},\cdots,b_{M-1}}{}
           \phi_{0}^{(0)}(b_{0}|u,v,w)\phi_{1}^{(0)}(b_{1}|u,v,w)
           \cdots \phi_{M-1}^{(0)}(b_{M-1}|u,v,w)
           {\cal L}(u,v,w)
           ^{b_{0}b_{1}\cdots b_{M-1}}_{a_{0}a_{1}\cdots a_{M-1}}
 & \nonumber \\
 &= \dprod{j=0}{M-1}\phi_{j}^{(1)}(a_{j}|u,v,w)
    +\dprod{j=0}{M-1}\phi_{j}^{(2)}(a_{j}|u,v,w).&
 \label{eq:prevecFR}
\end{eqnarray}
Defining vectors $\psi^{(\nu)}(u,v,w)\in ({\bf C}^{N})^{\otimes M}$
by
\begin{equation}
 \psi^{(\nu)}(u,v,w)_{a_{0}a_{1}\cdots a_{M-1}}
 =\prod_{j=0}^{M-1}\phi_{j}^{(\nu)}(a_{j}|u,v,w),
 \label{eq:vector}
\end{equation}
we can write (\ref{eq:prevecFR}) as
\begin{equation}
 {\cal L}(u,v,w)\,\psi^{(0)}(u,v,w)
 =\psi^{(1)}(u,v,w)+\psi^{(2)}(u,v,w).
 \label{eq:vecFR}
\end{equation}
In the following, we will find
a family of solutions to (\ref{eq:phiP})
\begin{eqnarray*}
 &
 \phi_{j}^{(\nu)}(b|u,v,w)=\phi^{(\nu)}(c_{j},b,c_{j+1}|u,v,w)
 \quad(\nu=0,1,2,3\ \mbox{and}\,b\in{\bf Z}/N{\bf Z}),
 & \\
 &
 P_{j}=P(c_{j}),
 &
\end{eqnarray*}
labeled by
$\Bigl\{\,(c_{0},c_{1},\cdots,c_{M-1})\ |
    \ c_{j}\in{\bf Z}/N{\bf Z}\ \ \mbox{for}\ j\in{\bf Z}/M{\bf Z}\,\Bigr\}$.
This gives rise to $N^{M}$ vectors $\psi^{(\nu)}$
labeled as
$\psi^{(\nu)}(u,v,w)^{c_{0},c_{1},\cdots,c_{M-1}}$.
We will also prove that $\psi$'s are proportional to the row vectors
of the diagonal-to-diagonal transfer matrix $\Phi$
of the broken ${\bf Z}_{N}$-symmetric model,
\begin{eqnarray}
 &\psi^{(0)}(\lambda-u,\lambda-v,w+1/4)
 ^{c_{0}c_{1}\cdots c_{M-1}}_{b_{0}b_{1}\cdots b_{M-1}}
 =
 \Phi(u,v,w)^{c_{0}c_{1}\cdots c_{M-1}}_{b_{0}b_{1}\cdots b_{M-1}},&
 \label{eq:psi0}\\
 \lefteqn{
 \psi^{(1)}(\lambda-u,\lambda-v,w+1/4)
 ^{c_{0}c_{1}\cdots c_{M-1}}_{b_{0}b_{1}\cdots b_{M-1}}}\nonumber\\
 &=
 \Bigl(C(u,v,w)f(u,v,w)\Bigr)^{M}
 \Phi(u,v,w+\eta)
 ^{c_{0}c_{1}\cdots c_{M-1}}_{b_{0}b_{1}\cdots b_{M-1}},&
 \label{eq:psi1}\\
 \lefteqn{
 \psi^{(2)}(\lambda-u,\lambda-\lambda-v,w+1/4)
 ^{c_{0}c_{1}\cdots c_{M-1}}_{b_{0}b_{1}\cdots b_{M-1}}}\nonumber\\
 &=
 \Bigl(-C(u,v,w)f(\lambda-v,\lambda-u,-w)\Bigr)^{M}
 \Phi(u,v,w-\eta)
 ^{c_{0}c_{1}\cdots c_{M-1}}_{b_{0}b_{1}\cdots b_{M-1}}.&
 \label{eq:psi2}
\end{eqnarray}
The results (\ref{eq:vecFR}) to (\ref{eq:psi2})
altogether implies the functional relation (\ref{eq:matFR}).

Now we start to solve (\ref{eq:phiP}).
We write the matrix elements of $P_{j}$  as
\begin{displaymath}
 P_{j}=\left(\begin{array}{cc}
               p_{j}^{(0)} & p_{j}^{(2)} \\ p_{j}^{(1)} & p_{j}^{(3)}
   \end{array}\right),
\end{displaymath}
and its first column vector $\mbox{}^{t}(p_{j}^{(0)},p_{j}^{(1)})$
as ${\bf p}_{j}$.
Multiplying $P_{j}$ to (\ref{eq:phiP}) from the left
and taking its first column, we have
\begin{equation}
 \sum_{b=0}^{N-1}\phi_{j}^{(0)}(b|u,v,w)L(a,b|u,v,w)\,
 {\bf p}_{j+1}
 =\phi_{j}^{(1)}(a|u,v,w)\,{\bf p}_{j}
 \quad
 \mbox{for}\quad a\in{\bf Z}/N{\bf Z}.
 \label{eq:defphi}
\end{equation}
For later use, we define the functions
$\Delta_{\ast(\pm)}$, $\Delta^{\ast}_{(\pm)}$,
$\delta_{\ast}$ and $\delta^{\ast}$ by
\begin{eqnarray*}
 &
 \Delta_{\ast(\pm)}\Bigl({\bf p},a\,\Big|u\Bigl)
 = p^{(0)}\,K\mbox{}^{\ a\pm 1}_{1 \ \,a}(u)
  -p^{(1)}\,K\mbox{}^{\ a\pm 1}_{0 \ \,a}(u),
 & \\
 &
 \Delta^{\ast}_{(\pm)}\Bigl(a,{\bf p}\,\Big|u\Bigl)
 = K\mbox{}^{0a\pm 1}_{\ \ \,a}(u)\,p^{(0)}
  +K\mbox{}^{1a\pm 1}_{\ \ \,a}(u)\,p^{(1)},
 & \\
 &
 \delta_{\ast}(a|u)
 = K\mbox{}^{\ a-1}_{0\ \,a}(u)\,K\mbox{}^{\ a+1}_{1\ \,a}(u)
  -K\mbox{}^{\ a+1}_{0\ \,a}(u)\,K\mbox{}^{\ a-1}_{1\ \,a}(u),
 & \\
 &
 \delta^{\ast}(a|u)
 = K\mbox{}^{0a-1}_{\ \ \,a}(u)\,K\mbox{}^{1a+1}_{\ \ \,a}(u)
  -K\mbox{}^{0a+1}_{\ \ \,a}(u)\,K\mbox{}^{1a-1}_{\ \ \,a}(u).
 &
\end{eqnarray*}
The equations (\ref{eq:defphi}) constitute a system of $2N$
homogeneous linear equations
in $\phi_{j}^{(0)}(a|u,v,w)$ and $\phi_{j}^{(1)}(a|u,v,w)$
for $a$ $\in{\bf Z}/N{\bf Z}$.
It has a non-trivial solution if and only if the determinant
of its coefficient matrix vanishes.
Demanding this condition, we obtain
\begin{eqnarray}
 \lefteqn{
 \dprod{a=0}{N-1}
 \Delta_{\ast(-)}({\bf p}_{j},a|u-w)\,
 \Delta^{\ast}_{(-)}(a,{\bf p}_{j+1}|v-w)} \nonumber \\
 &+&
 \dprod{a=0}{N-1}
 \Delta_{\ast(+)}({\bf p}_{j},a|u-w)\,
 \Delta^{\ast}_{(+)}(a,{\bf p}_{j+1}|v-w)
 =0.
 \label{eq:detcond}
\end{eqnarray}
Later we find that the equation (\ref{eq:detcond})
restricts ${\bf p}_{j}$ to a discrete set of values.
When we parameterize ${\bf p}_{j}$ as
\begin{equation}
 {\bf p}_{j}={\bf p}(c_{j}),\quad
 {\bf p}(c)
 =
 \left(\begin{array}{c} p^{(0)}(c) \\ p^{(1)}(c) \end{array}\right)
 =
 \left(\begin{array}{c}
       {[}\theta_{2}\theta_{3}{]}(c\eta) \\
      -{[}\theta_{1}\theta_{4}{]}(c\eta)\end{array}\right),
 \label{eq:parap}
\end{equation}
and denote the dependence on ${\bf p}(c)$ simply by $c$
and $U_{\alpha}=u+\alpha\eta$,
$\Delta$'s and $\delta$'s become
\begin{eqnarray}
  \Delta_{\ast(\pm)}(c,a|u-1/4)
  &= &
  \displaystyle{\frac{{[}\theta_{2}\theta_{3}{]}(0)}
                          {G_{a}G_{a\pm 1}}}\,
  \theta_{2}(U_{\pm c\mp a})\,
  \theta_{3}(U_{\mp c\mp a}),
  \nonumber \\
  \Delta^{\ast}_{(\pm)}(a,c|u-1/4)
  &= &
  {[}\theta_{2}\theta_{3}{]}(0)\,
      \theta_{1}(U_{\pm a\mp c+ 1})\,
      \theta_{4}(U_{\pm a\pm c+ 1}),
  \label{eq:deltas} \nonumber \\
  \delta_{\ast}(a|u-1/4)
  &= &
  -\displaystyle{\frac{{[}\theta_{2}\theta_{3}\theta_{4}{]}(0)}
         {G_{a-1}G_{a+1}}}\,\theta_{1}(2U_{0}),
  \nonumber \\
  \delta^{\ast}(a|u-1/4)
  &= &
  {[}\theta_{2}\theta_{3}\theta_{4}{]}(0)\,G_{a}^{2}\,
      \theta_{1}(2U_{1}).
  \nonumber
\end{eqnarray}
Under the parameterization (\ref{eq:parap}),
the condition (\ref{eq:detcond}) holds
if and only if either $c_{j}$'s
are all integers, or all half-integers. We restrict ourselves to the case
that $c_{j}$'s are all integers, because only in this case
the relation (\ref{eq:vecFR}) gives the functional relation
(\ref{eq:matFR}).
The system of equations (\ref{eq:defphi}) involves not all $\phi$'s
but only
$\phi_{j}^{(0)}(a+1|u,v,w)$, $\phi_{j}^{(0)}(a-1|u,v,w)$
and $\phi_{j}^{(1)}(a|u,v,w)$.
Expressing $\phi_{j}^{(0)}(a+1|u,v,w)$ and
$\phi_{j}^{(1)}(a|u,v,w)$
in terms of $\phi_{j}^{(0)}(a-1|u,v,w)$, we obtain
\begin{eqnarray}
 &
 \displaystyle{\frac{\phi_{j}^{(0)}(a+1|u,v,w)}
                    {\phi_{j}^{(0)}(a-1|u,v,w)}}
 =
 -\displaystyle{
                \dfrac{\Delta_{\ast(-)}(c_{j},a|u-w)\,
                       \Delta^{\ast}_{(-)}(a,c_{j+1}|v-w)}
                      {\Delta_{\ast(+)}(c_{j},a|u-w)\,
                       \Delta^{\ast}_{(+)}(a,c_{j+1}|v-w)}},
 & \label{eq:recphi0} \\
 &
 \displaystyle{\frac{\phi_{j}^{(1)}(a|u,v,w)}
                    {\phi_{j}^{(0)}(a-1|u,v,w)}}
 = \delta_{\ast}(a|u-w)
 \displaystyle{\frac{\Delta^{\ast}_{(-)}(a,c_{j+1}|v-w)}
                    {\Delta_{\ast(+)}(c_{j},a|u-w)}},
 & \label{eq:recphi1}
\end{eqnarray}
{}From (\ref{eq:recphi0}) and (\ref{eq:recphi1}), we can write
$\phi_{j}^{(\nu)}(a|u,v,w)$ ($\nu=0,1$) as
\begin{equation}
 \phi_{j}^{(\nu)}(a|u,v,w)=\phi^{(\nu)}(c_{j},a,c_{j+1}|u,v,w),
\end{equation}
where the function $\phi^{(\nu)}(c,a,c'|u,v,w)$ is independent of $j$.
Taking the determinant of both sides of (\ref{eq:phiP}), we find
\begin{eqnarray}
 &
 \delta_{\ast}(c_{j},a|u-w)\,\delta^{\ast}(a,c_{j+1}|v-w)\,
          \displaystyle{\frac{\mbox{det}(P_{j+1})}{\mbox{det}(P_{j})}}
 & \nonumber \\
 &= \dfrac{\phi^{(1)}(c_{j},a,c_{j+1}|u,v,w)\,
           \phi_{j}^{(2)}(a|u,v,w)}
          {\phi^{(0)}(c_{j},a-1,c_{j+1}|u,v,w)\,
           \phi^{(0)}(c_{j},a+1,c_{j+1}|u,v,w)}.
 & \label{eq:prerecphi2}
\end{eqnarray}
We set det$(P_{j})$ to unity without loss of generality.
Then the equations (\ref{eq:recphi0}), (\ref{eq:recphi1})
and (\ref{eq:prerecphi2}) give
\begin{equation}
 \displaystyle{\frac{\phi^{(2)}(c_{j},a,c_{j+1}|u,v,w)}
                    {\phi^{(0)}(c_{j},a-1,c_{j+1}|u,v,w)}}
 = -\delta^{\ast}(a|v-w)
 \displaystyle{\frac{\Delta_{\ast(-)}(c_{j},a|u-w)}
                    {\Delta^{\ast}_{(+)}(a,c_{j+1}|v-w)}},
 \label{eq:recphi2}
\end{equation}
where we write $\phi_{j}^{(2)}(a|u,v,w)$ as
$\phi^{(2)}(c_{j},a,c_{j+1}|u,v,w)$.
The relations
(\ref{eq:recphi0}), (\ref{eq:recphi1}) and (\ref{eq:recphi2})
recursively determine $\phi^{(\nu)}_{a}$'s.
We abbreviate $u-w+\alpha\eta$ and $v-w+\gamma\eta$
to $A_{\alpha}$ and $B_{\gamma}$ respectively.
Comparing (\ref{eq:recphi0}) with (\ref{eq:recW}) and (\ref{eq:recWb}),
we have
\begin{equation}
 \displaystyle{\frac{\phi^{(0)}(a,b+1,c|u,v,w+1/4)}
                    {\phi^{(0)}(a,b-1,c|u,v,w+1/4)}}
 =
 \displaystyle{\frac{W(a,b+1|A_{0})\,\overline{W}(b+1,c|B_{0})}
                    {W(a,b-1|A_{0})\,\overline{W}(b-1,c|B_{0})}}.
 \label{eq:phi0W}
\end{equation}
Hence we find that $\phi^{(0)}$ is a product of the two
Boltzmann weights,
\begin{equation}
 \phi^{(0)}(a,b,c|u,v,w+1/4)
 =W(a,b|A_{0})\,\overline{W}(b,c|B_{0}).
 \label{eq:finalphi0}
\end{equation}
{}From (\ref{eq:recphi0}) and (\ref{eq:recphi1}),
we obtain
\begin{equation}
 \displaystyle{\frac{\phi^{(1)}(a,b+1,c|u,v,w+1/4)}
                    {\phi^{(1)}(a,b-1,c|u,v,w+1/4)}}
 =
  \displaystyle{\frac{\phi^{(0)}(a,b+1,c|u-\eta,v-\eta,w+1/4)}
                     {\phi^{(0)}(a,b-1,c|u-\eta,v-\eta,w+1/4)}}.
 \label{eq:phi1phi0}
\end{equation}
The same procedure for $\phi^{(2)}$ yields
\begin{equation}
 \displaystyle{\frac{\phi^{(2)}(a,b+1,c|u,v,w+1/4)}
                    {\phi^{(2)}(a,b-1,c|u,v,w+1/4)}}
 =
  \displaystyle{\frac{\phi^{(0)}(a,b+1,c|u+\eta,v+\eta,w+1/4)}
                    {\phi^{(0)}(a,b-1,c|u+\eta,v+\eta,w+1/4)}}
 \label{eq:phi2phi0}
\end{equation}
By (\ref{eq:finalphi0}), (\ref{eq:phi1phi0}) and
(\ref{eq:phi2phi0}), we can write $\phi^{(1)}$ and $\phi^{(2)}$ as
\begin{eqnarray}
 \phi^{(1)}(a,b,c|u,v,w+1/4)
 &=&
 f_{ac}(u,v,w)\,W(a,b|A_{-1})\,\overline{W}(b,c|B_{-1}),
 \label{eq:phi1W} \\
 \phi^{(2)}(a,b,c|u,v,w+1/4)
 &=&
 g_{ac}(u,v,w)\,W(a,b|A_{1})\,\overline{W}(b,c|B_{1}),
 \label{eq:phi2W}
\end{eqnarray}
where $f_{ac}$ and $g_{ac}$ are functions independent of $b$.
The equations (\ref{eq:recphi1}), (\ref{eq:finalphi0})
and (\ref{eq:phi1W}) determine $f_{ac}(u,v,w)$ as
\begin{eqnarray*}
 \lefteqn{f_{ac}(u,v,w)} \nonumber \\
 &=& {[}\theta_{2}\theta_{3}\theta_{4}{]}(0)\,\theta_{1}(2A_{0})\,
     \displaystyle{\frac{G_{b}}{G_{b-1}}}
     \displaystyle{\frac{\theta_{1}(B_{-b+c+1})\,\theta_{4}(B_{-b-c+1})}
                        {\theta_{2}(A_{a-b})\,\theta_{3}(A_{-a-b})}}
     \displaystyle{\frac{W(a,b-1|A_{0})\,\overline{W}(b-1,c|B_{0})}
                        {W(a,b|A_{-1})\,\overline{W}(b,c|B_{-1})}}
     \nonumber \\
 &=& C(u,v,w)\,\theta_{1}(2A_{0})\,
     \displaystyle{\frac{{[}\theta_{1}\theta_{4}{]}(B_{1})}
                        {{[}\theta_{2}\theta_{3}{]}(A_{0})}}
     =C(u,v,w)\,f(u,v,w),
 \label{eq:f}
\end{eqnarray*}
where $C(u,v,w)$ and $f(u,v,w)$ were given
in (\ref{eq:C}) and (\ref{eq:newf}), respectively.
The last equality is due to (\ref{eq:identity1}).
In the same way, we obtain
\begin{displaymath}
 g_{ac}(u,v,w)
 =
 -C(u,v,w)\,f(\lambda-v,\lambda-u,-w).
 \label{eq:g}
\end{displaymath}
The equations (\ref{eq:phi1W}) and (\ref{eq:phi2W}) become
\begin{equation}
 \phi^{(1)}(a,b,c|u,v,w+1/4)
 =
 C(u,v,w)\,f(u,v,w)\,W(a,b|A_{-1})\,\overline{W}(b,c|B_{-1}),
 \label{eq:finalphi1}
\end{equation}
\begin{equation}
 \phi^{(2)}(a,b,c|u,v,w+1/4)
 =
 -C(u,v,w)\,f(\lambda-v,\lambda-u,-w)\,
 W(a,b|A_{1})\,\overline{W}(b,c|B_{1}).
 \label{eq:finalphi2}
\end{equation}
Substituting (\ref{eq:finalphi0}), (\ref{eq:finalphi1})
and (\ref{eq:finalphi2}) into the definition (\ref{eq:vector})
of $\psi^{(\nu)}(u,v,w)$ and using the crossing symmetry (\ref{eq:CroSym}),
we obtain (\ref{eq:psi0}), (\ref{eq:psi1}) and (\ref{eq:psi2}).
We have the functional relation (\ref{eq:matFR}) as a result.

\section{Bethe Ansatz Equations}
\setcounter{equation}{0}
\label{sec:BAE}
In this section, we give commutation relations
among ${\cal R}$, ${\cal L}(u)$ and $\Phi(v)$, and reduce
the functional relation (\ref{eq:matFR})
to the functional equation among their eigenvalues.
After discussing some properties about the zeros and poles of the eigenvalues
of $\Phi(u)$, we derive the Bethe Ansatz equations
(\ref{eq:introBAE}), (\ref{eq:introuv}) and (\ref{eq:sumcond}) for the
broken ${\bf Z}_{N}$-symmetric model.

First we have
\begin{equation}
 {\cal L}(u,v,w')\,\Phi(u,v,w)
 = \Phi(u,v,w)\,{\cal L}(v,u,w').
 \label{eq:LPHI1}
\end{equation}
We give a proof in Appendix \ref{ap:ComRel}.
In the case of the homogeneous system, i.e., $u=v$ and $w=w'$,
the equation (\ref{eq:LPHI1}) means the commutativity of two
transfer matrices ${\cal L}(u)={\cal L}(u,u,w)$
and $\Phi(u)=\Phi(u,u,w)$,
\begin{equation}
 \Bigl[\;{\cal L}(u)\; ,\;\Phi(v)\;\Bigr]=0.
 \label{eq:LPHI2}
\end{equation}
The star-triangle relation (\ref{eq:STR}) gives
\begin{equation}
 \Bigl[\;\Phi(u)\; ,\;\Phi(v)\;\Bigr]=0,
 \label{eq:PHIPHI}
\end{equation}
and the $LLR=RLL$ relation (\ref{eq:LLR}) guarantees
\begin{equation}
 \Bigl[\;{\cal L}(u)\; ,\;{\cal L}(v)\;\Bigr]=0.
 \label{eq:LL}
\end{equation}
The relations ({\ref{eq:LPHI2}), ({\ref{eq:PHIPHI}) and ({\ref{eq:LL})
make it possible to diagonalize ${\cal L}(u)$ and $\Phi(v)$
simultaneously by eigenvectors independent of the spectral
parameters $u$ and $v$.
Fixing one of the eigenvectors and denoting the corresponding
eigenvalues of ${\cal L}(u)$ and $\Phi(v)$ by
$l(u)$ and $\varphi(v)$ respectively, we can rewrite
the functional relation (\ref{eq:matFR}) as
\begin{equation}
 l(\lambda-u-1/4)\,\varphi(u)
 =
 C(u)^{M}\Biggl(f(u)^{M}\varphi(u-\eta)
               +(-1)^{M}f(\lambda-u)^{M}\varphi(u+\eta)\Biggr),
 \label{eq:eigenFR}
\end{equation}
where from (\ref{eq:newf}) and (\ref{eq:C}), $f(u)$ and $C(u)$ are
\begin{equation}
 f(u)=\theta_{1}(2u)\,\dfrac{{[}\theta_{1}\theta_{4}{]}(u+\eta)}
                            {{[}\theta_{2}\theta_{3}{]}(u)}
 \quad \mbox{and} \quad
 C(u)={[}\theta_{2}\theta_{3}\theta_{4}{]}(0)
      \,{[}T_{2}^{(+)}T_{3}^{(+)}T_{2}^{(-)}T_{3}^{(-)}{]}(n|u).
 \label{eq:fandC}
\end{equation}
The next step to derive the Bethe Ansatz equations
is to examine the quasi-periodicity property of $\varphi(u)$.
We have the following relations
\begin{eqnarray}
 &\Phi(u+1) = \Phi(u),&
 \label{eq:PHIperiod} \\
 &{\cal R}\,\Phi(u) = \Phi(u)\,{\cal R}
                    = \Phi(u+\dfrac{\tau}{2}).&
 \label{eq:RPHI}
\end{eqnarray}
We have defined ${\cal R}$ in (\ref{eq:defR}).
The periodicity (\ref{eq:PHIperiod}) is obvious
from the definition (\ref{eq:defPHI}) of $\Phi$
and the periodicity (\ref{eq:Wperiod1}) of $W$ and $\overline{W}$.
The other periodicity (\ref{eq:RPHI}) follows from (\ref{eq:WWbperiod}).
We also have the commutativity
\begin{equation}
 \Bigl[\;{\cal R}\; ,\;{\cal L}(u)\;\Bigr]=0,
\end{equation}
from the ${\bf Z}_{2}$-symmetry of $K$'s (\ref{eq:Z2K})
and the definitions (\ref{eq:defR}) and (\ref{eq:defL}) .
Diagonalizing ${\cal R}$ and $\Phi(u)$ simultaneously
with ${\cal L}(v)$,
the equations
(\ref{eq:PHIperiod}), (\ref{eq:RPHI}) and (\ref{eq:defR})
give
\begin{equation}
 \varphi(u+1) = \varphi(u),
 \quad
 \varphi(u+\dfrac{\tau}{2}) = r\,\varphi(u),
 \quad r=\pm 1,
\label{eq:phiperiod}
\end{equation}
where $r$ is an eigenvalue of ${\cal R}$.
The poles of $\varphi(u)$ are coming from
only those of the matrix elements of $\Phi(u)$.
We define
\begin{eqnarray}
 p(u) &=& \biggl(\dfrac{\eta(\tau)}{\eta(2\tau)^{2}}\biggr)^{n}
          \prod_{j=1}^{n}{[}\theta_{2}\theta_{3}{]}
                         (u-(2j-1)\eta|\tau) \nonumber\\
      &=& \prod_{j=1}^{n}\theta_{2}(u-(2j-1)\eta|\tau/2) ,
 \label{eq:p}
\end{eqnarray}
which contains all possible poles of $W(a,b|u)$.
The same does $p(\lambda-u)$ for $\overline{W}$
by the crossing symmetry.
Hence the set of zeros
of $\Bigl(p(u)p(\lambda-u)\Bigr)^{M}$
contains all poles of $\varphi(u)$.
By the Lemma in Appendix \ref{ap:theta}
and the double periodicity (\ref{eq:phiperiod})
of $\varphi(u)$,
we can write $\varphi(u)$ as
\begin{eqnarray*}
 &
 \varphi(u)=\bigl(\mbox{{\it const}})\,
            \dfrac{\displaystyle{\prod_{j=1}^{2nM}}
                   \theta_{1}(u-u_{j}|\tau/2)}
                  {\biggl(p(u)\,p(\lambda-u)\biggr)^{M}},
 & \label{eq:eigenvalue} \\
 &
 \displaystyle{\sum_{j=1}^{2nM}}u_{j} \equiv nM\lambda +
 \dfrac{1-r}{4}\ \bmod\bigl({\bf Z} \oplus \dfrac{\tau}{2}{\bf Z}).
 &
\end{eqnarray*}
The initial condition $\Phi(0)=Id$ (\ref{eq:WWbzero})
determines the normalization of $\mbox{{\it const}}$,
\begin{equation}
 \varphi(u)=\Biggl(
            \dfrac{p(0)p(\lambda)}{p(u)p(\lambda-u)}
            \Biggr)^{M}\,
            \dprod{j=1}{2nM}
            \dfrac{\theta_{1}(u-u_{j}|\tau/2)}
                  {\theta_{1}(u_{j}|\tau/2)}.
 \label{eq:eigenvalue2}
\end{equation}
Assuming $C(u_{j})\ne 0$ in (\ref{eq:fandC}) and
substituting $u_{k}$ ($k=1,\cdots,2nM$) into (\ref{eq:eigenFR}),
we have
\begin{eqnarray}
 &
 f(u_{k})^{M}\varphi(u_{k}-\eta)
 +(-1)^{M}f(\lambda-u_{k})^{M}\varphi(u_{k}+\eta)=0
 & \nonumber \samepage \\
 &
 \hspace{70mm}\mbox{for}\quad k=1,\cdots,2nM.
 &
 \label{eq:temptempBAE}
\end{eqnarray}
We further assume that $u_{k}$ ($k=1,\cdots,2nM$)
are neither zeros nor poles of $f(u)$, $f(\lambda-u)$
and $\varphi(u\pm\eta)$.
Then (\ref{eq:temptempBAE}) becomes
\begin{eqnarray}
 &
 \Biggl(
 \dfrac{f(u_{k})}{f(\lambda-u_{k})}
 \dfrac{p(u_{k}+\eta)\,p(\lambda-u_{k}-\eta)}
       {p(u_{k}-\eta)\,p(\lambda-u_{k}+\eta)}
 \Biggr)^{M}
  =
 (-1)^{M+1}\displaystyle{\prod_{j=1}^{2nM}}
 \dfrac{\theta_{1}(u_{k}-u_{j}+\eta|\tau/2)}
       {\theta_{1}(u_{k}-u_{j}-\eta|\tau/2)}
 & \nonumber \\
 &  \hfill \hspace{70mm}
 \mbox{for}\quad k=1,\cdots,2nM.
 &
 \label{eq:tempBAE}
\end{eqnarray}
We can write $f(u)$ in (\ref{eq:fandC}) by (\ref{eq:tautau}) as
\begin{displaymath}
 f(u)=\dfrac{\eta(2\tau)}{\eta(\tau)^{2}}\,
      \theta_{1}(u|\tau/2)\,
      \theta_{1}(u+\eta|\tau/2).
\end{displaymath}
Then the left-hand side of (\ref{eq:tempBAE}) reduces to
$\Biggl(\dfrac{\theta_{1}(u_{k}|\tau/2)}
               {\theta_{1}(u_{k}-\lambda|\tau/2)}\Biggr)^{2M}$.
After shifting $u_{k}$ by $\lambda/2$, i.e., putting
\begin{displaymath}
 v_{k}=u_{k}-\dfrac{\lambda}{2}
 \quad \mbox{for}\quad k=1,\cdots,2nM,
\end{displaymath}
we obtain
the Bethe Ansatz equations for the broken ${\bf Z}_{N}$-symmetric
model,
\begin{eqnarray}
 &
 \Biggl(
 \dfrac{\theta_{1}(v_{k}+\lambda/2|\tau/2)}
       {\theta_{1}(v_{k}-\lambda/2|\tau/2)}
 \Biggr)^{2M}
 =(-1)^{M+1}
 \displaystyle{\prod_{j=1}^{2nM}}
 \dfrac{\theta_{1}(v_{k}-v_{j}+\eta|\tau/2)}
       {\theta_{1}(v_{k}-v_{j}-\eta|\tau/2)}
 &\nonumber \\
 &
 \hspace{70mm}\mbox{for}\quad k=1,\cdots,2nM,
 & \label{eq:BAE} \\
 &
 \displaystyle{\sum_{j=1}^{2nM}}v_{j} \equiv
 \dfrac{1-r}{4}\  \bmod\bigl({\bf Z} \oplus \dfrac{\tau}{2}{\bf Z}).
 & \label{eq:condition}
\end{eqnarray}

\section{Density Function and Free Energy}
\setcounter{equation}{0}
\label{sec:energy}
In this section, we will calculate the free energy of
the broken ${\bf Z}_{N}$-symmetric model from the Bethe Ansatz
equations under the three assumptions concerning the ground state.
One is the String Hypothesis below, and the others are about
the distribution of string centers
$v_{\alpha}=\sqrt{-1}w_{\alpha}$ and the corresponding
quantum numbers $I_{\alpha}$.
We restrict the spectral parameter $u$
to the region ${[}0,1/2N{]}$, in which
all the Boltzmann weights are real and positive,
and $\tau$ to a pure imaginary number, $\tau=\sqrt{-1}\kappa$
with $\kappa$ real and positive.

By a string of length $l$ and parity $\nu$ ($=$ $0$ or $1$) with
its center $v_{\alpha}$, we mean the following set,
\begin{equation}
 \left\{ \begin{array}{cc}
         v_{\alpha,j} &
         \left|\begin{array}{c}
            v_{\alpha,j} \equiv v_{\alpha}+(2j-l-1)
            \dfrac{\eta}{2}+\dfrac{\nu}{2} \quad
            \bmod({\bf Z} \oplus \dfrac{\tau}{2}{\bf Z})\\
            \qquad\mbox{for}\quad j=1,2,\cdots,l,
            \quad \mbox{and}\quad v_{\alpha}:\, \mbox{pure imaginary}
         \end{array}\right.
 \end{array}\right\}.
 \label{string}
\end{equation}
We suppose that the following hypothesis holds in the infinite lattice limit
\cite{Albertini92}\cite{ADMc91}\cite{TS72}. \\
\hspace*{20mm}\underline{String Hypothesis for the ground state}

\vspace{3mm}

\mbox{\ \ \ \ \ \ \ \ \ \ \ \
  }\begin{minipage}{11cm}
  {\it
  The solution of the BAE's (\ref{eq:BAE}), $\{v_{j}, j=1,\cdots,2nM\}$,
  corresponding to the ground state consists of
  strings of length $N-1$ and parity \quad
  $\dfrac{1-(-1)^{n+1}}{2}$.
  }
  \end{minipage}

\vspace{7mm}

More precisely, for finite systems the
solutions of the BAE's may have deviations
from strings.
The hypothesis asserts that these deviations vanish
in the infinite lattice limit.
In the course of the following calculation
we deal with the solutions of the BAE's as if they
were genuine strings,
since we are interested in thermodynamic quantities.
Because all the matrix elements of $\Phi$ are real and positive,
the Perron-Frobenius theorem \cite{Gantmacher59} shows that the ground state
belongs to the sector of zero quasi-momentum \cite{Albertini92}.
The hypothesis implies that the ground state also belongs
to the sector $r=1$, and that
the corresponding solutions are made up
of $M$ strings of length $2n$.
We denote them by
\begin{eqnarray*}
 &
 v_{\alpha,j} \equiv \sqrt{-1} w_{\alpha}+(N-2j)
 \dfrac{\eta}{2}+\dfrac{1-(-1)^{n+1}}{2}\quad
 \bmod({\bf Z} \oplus \dfrac{\tau}{2}{\bf Z})
 & \nonumber \\
 &
 \hspace{50mm}\mbox{for}\quad \alpha=1,\cdots,M
 \quad\mbox{and}\quad  j=1,\cdots,2n,
 &
\end{eqnarray*}
where $w_{\alpha}$'s are all real and taken as
\begin{displaymath}
 -\dfrac{\kappa}{4}\leq w_{1} \leq w_{2} \leq \cdots
 \leq w_{M} < \dfrac{\kappa}{4}.
\end{displaymath}
Then the BAE's for the ground state becomes
\begin{eqnarray}
 &
 \Biggl(
 \dfrac{\theta_{1}(v_{\beta,k}+\lambda/2|\tau/2)}
       {\theta_{1}(v_{\beta,k}-\lambda/2|\tau/2)}
 \Biggr)^{2M}
 =(-1)^{M+1}
 \displaystyle{\prod_{\alpha=1}^{M}}
 \displaystyle{\prod_{j=1}^{2n}}
 \dfrac{\theta_{1}(v_{\beta,k}-v_{\alpha,j}+\eta|\tau/2)}
       {\theta_{1}(v_{\beta,k}-v_{\alpha,j}-\eta|\tau/2)}
 &\nonumber \\
 &
 \hspace{50mm}\mbox{for}\quad \beta=1,\cdots,M,
 \quad k=1,\cdots,2n.
 & \label{eq:BAEGS}
\end{eqnarray}
Multiplying (\ref{eq:BAEGS}) over $ k=1,\cdots,2n$, we have
\begin{equation}
 \dfrac{\Biggl(
        \displaystyle{\prod_{k=1}^{2n}}
        \chi\Bigl(w_{\beta},\dfrac{n}{2N}(n-2k+\dfrac{1}{2n})
               +\dfrac{1-(-1)^{n+1}}{4}\Bigr)
        \Biggr)^{2M}}
       {\displaystyle{\prod_{\alpha=1}^{M}}
        \Biggl(
        \chi\Bigl(w_{\beta}-w_{\alpha},0\Bigr)
        \Bigl(\displaystyle{\prod_{j=1}^{2n-1}}
        \chi\Bigl(w_{\beta}-w_{\alpha},\dfrac{n}{N}j\Bigr)\Bigr)^{2}
        \chi\Bigl(w_{\beta}-w_{\alpha},\dfrac{2n^{2}}{N}\Bigr)
        \Biggr)}
 =1,
 \label{eq:BAEGS2}
\end{equation}
where $\chi(w,a)$ is
\begin{displaymath}
 \chi(w,a)=\dfrac{\theta_{1}(a-\sqrt{-1}w|\tau/2)}
                   {\theta_{1}(a+\sqrt{-1}w|\tau/2)}.
\end{displaymath}
Taking the logarithm of (\ref{eq:BAEGS2}) and
dividing it by $2\sqrt{-1}\pi M$, we have
\begin{equation}
 {\cal T}(w_{\beta})=\dfrac{I_{\beta}}{M}
 \quad \mbox{for}\quad \beta=1,\cdots,M,
 \label{eq:logBAE}
\end{equation}
where the quantum number $I_{\beta}$'s are integers and
\begin{eqnarray*}
 {\cal T}(w)
 &=&
 {\cal T}_{1}(w)
 -\dfrac{1}{2M}\displaystyle{\sum_{\alpha=1}^{M}}
 {\cal T}_{2}(w-w_{\alpha}),\\
 {\cal T}_{1}(w)
 &=&
 \displaystyle{\sum_{k=1}^{2n}}
 t\Bigl(w,\dfrac{n}{2N}(n-2k+\dfrac{1}{2n})+\dfrac{1-(-1)^{n+1}}{4}\Bigr),\\
 {\cal T}_{2}(w)
 &=&
 t(w,0)+2\displaystyle{\sum_{j=1}^{2n}}t(w,\dfrac{n}{N}j)
 -t(w,\dfrac{2n^{2}}{N}),\\
 t(w,a)
 &=&
 \dfrac{1}{\sqrt{-1}\pi} \log \chi(w,a).
\end{eqnarray*}
Albertini et al. numerically investigated the 3-state Fateev-Zamolodchikov
model \cite{ADMc91}.
Their results indicate that the String Hypothesis holds.
We assume that the centers $w$ of the strings are distributed densely
on the interval ${[}-\kappa/4,\kappa/4{]}$ in the limit of $M$ large,
and that the quantum numbers $I_{\beta}$ satisfy
\begin{equation}
 I_{\beta+1}=I_{\beta}+1\quad\mbox{for}\quad \beta=1,2,\cdots,2nM.
\end{equation}
These are the second and the third assumptions we make.
The results by Albertini et al. also supports them.
We furthermore conjecture that
\begin{equation}
 w_{\alpha}=-w_{M-\alpha+1}
 \label{eq:crossofw}
\end{equation}
holds exactly for the ground state even in the finite lattice.
This conjecture is consistent with their results.
If (\ref{eq:crossofw}) is true, we can show that
\begin{equation}
 {\cal T}(\kappa/4)-{\cal T}(-\kappa/4)=1,
 \label{eq:valuesofcalT}
\end{equation}
and this implies that $2M$ integers $I_{\beta}$'s must fill the interval
${[}-M,M{)}$ without jumps if all $I_{\beta}$'s are different.
This also supports our assumption.
But we do not use the conjecture (\ref{eq:crossofw}) in this Paper.

We now  proceed to the calculation.
We define the density function for $w$'s by
\begin{equation}
 \rho(w_{\beta})=\lim_{M\rightarrow \infty}
                 \dfrac{1}{M(w_{\beta+1}-w_{\beta})},
\end{equation}
which is positive and for any integrable function $f(x)$,
\begin{equation}
 \lim_{M\rightarrow \infty}\dfrac{1}{M}
 \sum_{\alpha=1}^{M}f(w_{\alpha})
 =
 \dint{-\frac{\kappa}{4}}{\frac{\kappa}{4}}
 f(w)\rho(w)\,dw
\end{equation}
holds.
Considering the difference of (\ref{eq:logBAE})
for $I_{\beta+1}$ and $I_{\beta}$
\begin{displaymath}
 \dfrac{1}{M(w_{\beta+1}-w_{\beta})}
 =\dfrac{I_{\beta+1}-I_{\beta}}{M(w_{\beta+1}-w_{\beta})}
 =\dfrac{{\cal T}(w_{\beta+1})-{\cal T}(w_{\beta})}
        {w_{\beta+1}-w_{\beta}},
\end{displaymath}
and letting $M\rightarrow \infty$, we obtain
\begin{equation}
 \rho(w)=\dfrac{d{\cal T}(w)}{dw}
        =\dfrac{d{\cal T}_{1}(w)}{dw}
         -\dfrac{1}{2} \int_{-\frac{\kappa}{4}}^{\frac{\kappa}{4}}
         \dfrac{d{\cal T}_{2}(w-\overline{w})}{dw}\rho(\overline{w})\,
         d\overline{w}.
 \label{eq:diffinteg}
\end{equation}
By (\ref{eq:conjugate}) and (\ref{eq:logexpan}),
we can expand $t(w,a)$ as
\begin{eqnarray*}
 t(w,a)
 &=&
 \dfrac{8\pi}{\kappa}\{a\}w+\dfrac{1}{\sqrt{-1}}\log
 \dfrac{\theta_{1}\Bigl(\dfrac{2\sqrt{-1}}{\kappa}\{a\}
                   +\dfrac{2}{\kappa}w\,\Big|-\dfrac{2}{\tau}\Bigr)}
       {\theta_{1}\Bigl(\dfrac{2\sqrt{-1}}{\kappa}\{a\}
                   -\dfrac{2}{\kappa}w\,\Big|-\dfrac{2}{\tau}\Bigr)}
 \nonumber \\
 &=&
 \dfrac{4\pi}{\kappa}(2\{a\}-1)+\sum_{k=1}^{\infty}
 \dfrac{\sin\Bigl(\dfrac{4\pi k}{\kappa}w\Bigr)\,
        \sinh\Bigl(\dfrac{4\pi k}{\kappa}(\{a\}-\dfrac{1}{2})\Bigr)}
       {k\,\sinh\Bigl(\dfrac{2\pi k}{\kappa}\Bigr)}.
\end{eqnarray*}
We denote the fractional part of $x$ by $\{x\}=x-{[}x{]}$,
${[}x{]}$ being the Gauss symbol.
When we write the Fourier expansions of $\dfrac{d{\cal T}_{1}(w)}{dw}$,
$\dfrac{d{\cal T}_{2}(w)}{dw}$ and $\rho(w)$ as
\begin{eqnarray*}
 \dfrac{d{\cal T}_{j}(w)}{dw}
 &=&
 \sum_{k=-\infty}^{\infty}A_{jk}\,
 \exp\Bigl(\dfrac{4\sqrt{-1}\pi k}{\kappa}w\Bigr)
 \quad\mbox{for}\quad j=1,2,\\
 \rho(w)
 &=&
 \sum_{k=-\infty}^{\infty}\rho_{k}\,
 \exp\Bigl(\dfrac{4\sqrt{-1}\pi k}{\kappa}w\Bigr),
\end{eqnarray*}
the integral equation (\ref{eq:diffinteg}) gives
\begin{displaymath}
 \rho_{k}=\dfrac{A_{1,k}}{1+\dfrac{\kappa}{4}A_{2,k}}.
\end{displaymath}
The coefficients $A_{jk}$ are
\begin{eqnarray*}
 A_{1,k}
 &=&
 \left\{\begin{array}{lc}
        \dfrac{4n}{N\kappa} & k=0, \\
        \dfrac{4\sinh\Bigl(\dfrac{\pi k}{N\kappa}(N-1)\Bigr)\,
               \cosh\Bigl(\dfrac{\pi k}{N\kappa}(N+1)\Bigr)}
              {\kappa\,\sinh\Bigl(\dfrac{2\pi k}{\kappa}\Bigr)\,
               \cosh\Bigl(\dfrac{2\pi k}{\kappa}\Bigr)}
                            & k \ne 0,
 \end{array} \right.  \\
 1+\dfrac{\kappa}{4}A_{2,k}
 &=&
 \left\{\begin{array}{lc}
        \dfrac{2n}{N} & k=0, \\
        \dfrac{2\sinh\Bigl(\dfrac{\pi k}{N\kappa}(N-1)\Bigr)\,
               \cosh\Bigl(\dfrac{\pi k}{N\kappa}(N+1)\Bigr)}
              {\sinh\Bigl(\dfrac{2\pi k}{\kappa}\Bigr)}
                            & k \ne 0.
 \end{array} \right.
\end{eqnarray*}
We obtain the density function for strings,
\begin{displaymath}
 \rho(w)
 =
 \dfrac{2}{\kappa}\sum_{k=-\infty}^{\infty}
 \dfrac{\exp\Bigl(\dfrac{4\sqrt{-1}\pi k}{\kappa}w\Bigr)}
       {\cosh\Bigl(\dfrac{\pi k}{N\kappa}\Bigr)}.
\end{displaymath}
Using (\ref{eq:conjugate}) and (\ref{eq:dnexpan}),
we can rewrite it as
\begin{eqnarray}
 \rho(w)
 &=&
 \dfrac{2}{\kappa}\,{[}\theta_{3}\theta_{4}{]}
                  \Bigl(0|-\dfrac{1}{N\tau}\Bigr)\,
 \dfrac{\theta_{3}\Bigl(\dfrac{2w}{\kappa}\,\Big|-\dfrac{1}{N\tau}\Bigr)}
       {\theta_{4}\Bigl(\dfrac{2w}{\kappa}\,\Big|-\dfrac{1}{N\tau}\Bigr)}
 \nonumber \\
 &=&
 2N\,{[}\theta_{2}\theta_{3}{]}\Bigl(0|N\tau\Bigr)\,
 \dfrac{\theta_{3}\Bigl(2\sqrt{-1}Nw\,|N\tau\Bigr)}
       {\theta_{2}\Bigl(2\sqrt{-1}Nw\,|N\tau\Bigr)}.
\end{eqnarray}
The free energy per site $F(u)$ of the model is defined by
\begin{equation}
 F(u)=\lim_{M\rightarrow\infty}
      \Biggl(-\dfrac{1}{M}\log\varphi(u)\Biggr),
 \label{eq:freeenergy}
\end{equation}
where $\varphi(u)$ is the eigenvalue of the transfer
matrix $\Phi(u)$ corresponding to the ground state.
We can write $\varphi(u)$ as
\begin{eqnarray*}
 &
 \varphi(u)
 =
 \Biggl(\dfrac{p(0)p(\lambda)}{p(u)p(\lambda-u)}\Biggr)^{M}
 \displaystyle{\prod_{j=1}^{2n}}
 \displaystyle{\prod_{\alpha=1}^{M}}D_{j}(u,w_{\alpha}),
 & \\
 &
 D_{j}(u,w)
 =
 \dfrac{\theta_{1}(\sqrt{-1}w+\beta_{j}-u|\tau/2)}
       {\theta_{1}(\sqrt{-1}w+\beta_{j}|\tau/2)},
 & \\
 &
 \beta_{j}=\{\gamma_{j}\},
 \quad
 \gamma_{j}=\dfrac{n}{2N}(N-2j+\dfrac{1}{2n})
                   +\dfrac{1-(-1)^{n+1}}{4},
 &
\end{eqnarray*}
by (\ref{eq:eigenvalue}), (\ref{eq:crossofw}) and
the String Hypothesis.
Then the free energy per site is
\begin{eqnarray*}
 F(u)
 &=&
 -\log\Biggl(\dfrac{p(0)p(\lambda)}{p(u)p(\lambda-u)}\Biggr)
 -\sum_{j=1}^{2n}\lim_{M\rightarrow\infty}
 \dfrac{1}{M}\sum_{\alpha=1}^{M}
 \log D_{j}(u,w_{\alpha}) \\
 &=&
 -\log\Biggl(\dfrac{p(0)p(\lambda)}{p(u)p(\lambda-u)}\Biggr)
 -\dfrac{1}{2}\sum_{j=1}^{2n}
 \int_{-\frac{\kappa}{4}}^{\frac{\kappa}{4}}
 \Bigl(\log D_{j}(u,w)\Bigr)\rho(w)\,dw.
\end{eqnarray*}
Since $\rho(w)$ is an even function, it is enough to integrate
the even part of $\log D_{j}(u,w)$.
We hence have
\begin{eqnarray*}
 F(u)
 &=&
 -\log\Biggl(\dfrac{p(0)p(\lambda)}{p(u)p(\lambda-u)}\Biggr) \\
 & &
 \qquad
 -\dfrac{1}{2}\sum_{j=1}^{2n}\log D^{(e)}_{j}(u,\dfrac{\kappa}{4})
 +\dfrac{1}{2}\sum_{j=1}^{2n}
 \int_{-\frac{\kappa}{4}}^{\frac{\kappa}{4}}
 \dfrac{d\log D^{(e)}_{j}(u,w)}{dw}\,\rho^{(-1)}(w)\,dw,
\end{eqnarray*}
where $D^{(e)}_{j}(u,w)$ and $\rho^{(-1)}(w)$ are
\begin{eqnarray*}
 &
 \rho^{(-1)}(w)=\dint{-\frac{\kappa}{4}}{w}
 \rho (\overl{w})\,d\overl{w},
 & \\
 &
 D^{(e)}_{j}(u,w)=\Bigl(D_{j}(u,w)D_{j}(u,-w)\Bigr)^{1/2}.
 &
\end{eqnarray*}
A little cumbersome calculation yields
\begin{eqnarray*}
 &
 -\log\Biggl(\dfrac{p(0)p(\lambda)}{p(u)p(\lambda-u)}\Biggr)
 =
 E_{1}(u)+E_{2}(u),
 & \\
 &
 -\dfrac{1}{2}\sum_{j=1}^{2n}\log D^{(e)}_{j}(u,\dfrac{\kappa}{4})
 =
 -E_{1}(u)+E_{3}(u),
 & \\
 &
 \dfrac{1}{2}\sum_{j=1}^{2n}
 \int_{-\frac{\kappa}{4}}^{\frac{\kappa}{4}}
 \dfrac{d\log D^{(e)}_{j}(u,w)}{dw}\,\rho^{(-1)}(w)\,dw
 =
 -E_{3}(u)+E_{4}(u),
 &
\end{eqnarray*}
where
\begin{displaymath}
 E_{1}(u)
 =
 \dfrac{2\pi n}{\kappa}\,u\Bigl(\dfrac{1}{N}-2u\Bigr),
\end{displaymath}
\begin{displaymath}
 E_{2}(u)
 =
 4\sum_{l=1}^{\infty}
 \dfrac{\sinh\Bigl(\dfrac{2\pi l}{\kappa}u\Bigr)\,
        \sinh\Bigl(\dfrac{2\pi l}{\kappa}(\dfrac{1}{2N}-u)\Bigr)\,
        \cosh\Bigl(\dfrac{\pi l}{\kappa}\Bigr)\,
        \sinh\Bigl(\dfrac{2\pi l}{N\kappa}n\Bigr)}
       {l\,\sinh\Bigl(\dfrac{2\pi l}{\kappa}\Bigr)\,
           \sinh\Bigl(\dfrac{2\pi l}{N\kappa}\Bigr)},
\end{displaymath}
\begin{displaymath}
 E_{3}(u)
 =
 -4\sum_{l=1}^{\infty}(-1)^{l}
 \dfrac{\sinh\Bigl(\dfrac{2\pi l}{\kappa}u\Bigr)\,
        \sinh\Bigl(\dfrac{2\pi l}{\kappa}(\dfrac{1}{2N}-u)\Bigr)\,
        \cosh\Bigl(\dfrac{2\pi l}{N\kappa}(n+1)\Bigr)\,
        \sinh\Bigl(\dfrac{2\pi l}{N\kappa}n\Bigr)}
       {l\,\sinh\Bigl(\dfrac{2\pi l}{\kappa}\Bigr)\,
           \sinh\Bigl(\dfrac{2\pi l}{N\kappa}\Bigr)},
\end{displaymath}
\begin{displaymath}
 E_{4}(u)
 =
 4\sum_{l=1}^{\infty}
 \dfrac{\sinh\Bigl(\dfrac{2\pi l}{\kappa}u\Bigr)\,
        \sinh\Bigl(\dfrac{2\pi l}{\kappa}(u-\dfrac{1}{2N})\Bigr)\,
        \cosh\Bigl(\dfrac{2\pi l}{N\kappa}(n+1)\Bigr)\,
        \sinh\Bigl(\dfrac{2\pi l}{N\kappa}n\Bigr)}
       {l\,\sinh\Bigl(\dfrac{2\pi l}{\kappa}\Bigr)\,
           \sinh\Bigl(\dfrac{2\pi l}{N\kappa}\Bigr)\,
           \cosh\Bigl(\dfrac{\pi l}{N\kappa}\Bigr)}.
\end{displaymath}
Now $F(u)$ has the final expression
\begin{eqnarray}
 F(u)
 &=&
 E_{2}(u)+E_{4}(u) \nonumber \\
 &=&
 -\sum_{l=1}^{\infty}
 \dfrac{\sinh\Bigl(\dfrac{2\pi l}{\kappa}u\Bigr)\,
        \sinh\Bigl(\dfrac{2\pi l}{\kappa}(\dfrac{1}{2N}-u)\Bigr)\,
        \sinh\Bigl(\dfrac{2\pi l}{N\kappa}n\Bigr)}
       {l\,\cosh\Bigl(\dfrac{\pi l}{\kappa}\Bigr)\,
           \cosh^{2}\Bigl(\dfrac{\pi l}{N\kappa}\Bigr)}.
 \label{eq:energy}
\end{eqnarray}
It agrees with the result of Jimbo, Miwa and Okado
\cite{JMO86} obtained by the use of the inversion-trick.
In the trigonometric limit
$\kappa\rightarrow +\infty$, the free energy formula (\ref{eq:energy})
reduces to the integral
\begin{displaymath}
 \lim_{\kappa\rightarrow\infty}F(u)
 =
 -\dint{0}{\infty}dx\,
 \dfrac{\sinh\Bigl(N\pi xu\Bigr)\,
        \sinh\Bigl(N\pi x(\dfrac{1}{2N}-u)\Bigr)\,
        \sinh\Bigl(n\pi x\Bigr)}
       {x\,\cosh\Bigl(\dfrac{N\pi x}{2}\Bigr)\,
           \cosh^{2}\Bigl(\dfrac{\pi x}{2}\Bigr)},
\end{displaymath}
which also agrees with the results
of Fateev and Zamolodchikov themselves \cite{FZ82}
and by Albertini \cite{Albertini92}.
The former was obtained by the inversion-trick, and the latter
by the Bethe Ansatz method.

\section{Discussion}
\setcounter{equation}{0}

The first main goal of this Paper is the functional relation
(\ref{eq:matFR}). We obtain it as a functional relation for
${\cal L}(u)$. The diagonal-to-diagonal transfer matrix $\Phi(u)$
of the broken  ${\bf Z}_{N}$-symmetric model appears naturally
in this relation.
We obtain the Boltzmann weights $W$ and $\overl{W}$
of the broken ${\bf Z}_{N}$-symmetric model in the
algebraic way different from that of \cite{HY90}.
We obtain $W$ and $\overl{W}$ as the solutions to the
relation (\ref{eq:defphi}). In \cite{HY90}, they are
the solutions to the relations (\ref{eq:defW}) and (\ref{eq:defWb}).

The Bethe Ansatz equations (\ref{eq:BAE}) are
the second goal of this Paper.
The commutativity (\ref{eq:LPHI1}) between ${\cal L}(u)$ and $\Phi(v)$
is essential to get the Bethe Ansatz equations (\ref{eq:BAE})
from the functional relation (\ref{eq:matFR}).
It is notable that the unitarity relations
(\ref{eq:unitarity1}) and (\ref{eq:unitarity2}) guarantee
this commutativity.
This contrasts with the usual situation where
the commutativity of the transfer matrices is
derived from the STR or the $LLR=RLL$ type relations.

The Fateev-Zamolodchikov model is the trigonometric limit
of the broken ${\bf Z}_{N}$-symmetric model.
It has the ${\bf Z}_{N}$-symmetry besides the ${\bf Z}_{2}$-symmetry.
Hence the ${\bf Z}_{N}$-charge $q$
$\in$ $\{-n,-n+1,\cdots,n-1,n\}$ is a good quantum number,
where $\exp(2\sqrt{-1}\pi\dfrac{q}{N})$ is an eigenvalue
of the ${\bf Z}_{N}$-charge operator
${\cal Q}$ $\in$ End($({\bf C}^{N})^{\otimes M}$),
\begin{eqnarray}
 &
 {\cal Q}=\overbrace{Q \otimes Q \otimes \cdots \otimes Q}
          ^{M\ times},
 & \nonumber \\
 &
 Q\,v_{j}^{(N)}
 = \exp\biggl(2\sqrt{-1}\pi\dfrac{j}{N}\biggr)v_{j}^{(N)}
 \quad\mbox{for}\quad j=0,1,\cdots,N-1.
 &
 \label{eq:defQ}
\end{eqnarray}
Albertini \cite{Albertini92} obtained
the Bethe Ansatz equations and the formula for the eigenvalue
$\varphi_{FZ}(u)$ of the diagonal-to-diagonal transfer matrix
for the Fateev-Zamolodchikov model.
They are
\begin{eqnarray}
 &
 \Biggl(
 \dfrac{\mbox{s}(v_{k}+\lambda/2)}
       {\mbox{s}(v_{k}-\lambda/2)}
 \Biggr)^{2M}
 =(-1)^{M+1}
 \displaystyle{\prod_{j=1}^{2nM-2|q|}}
 \dfrac{\mbox{s}(v_{k}-v_{j}+\eta)}
       {\mbox{s}(v_{k}-v_{j}-\eta)}
 \quad \mbox{for}\ k=1,\cdots,2nM,
 & \label{eq:AlbertiniBAE} \\
 &
 \varphi_{FZ}(u)
 =\Biggl(\dfrac{p_{\infty}(0)p_{\infty}(\lambda)}
               {p_{\infty}(u)p_{\infty}(\lambda-u)}\Biggr)^{\kern-0.3em M}
 \dprod{j=1}{\ 2nM-2|q|}\dfrac{s(u-u_{j})}{s(u_{j})},
 \quad
 p_{\infty}(u)=\lim_{\kappa\rightarrow\infty}p(u),
 & \label{eq:Albertinivarphi}
\end{eqnarray}
where $\mbox{s}(u)=\sin(\pi u)$ and $p(u)$ is given in (\ref{eq:p}).
The notations are slightly changed from \cite{Albertini92}
to compare to our case.
The main difference is the number of factors
in the right-hand sides of the BAE's and in the expressions
of the eigenvalues $\varphi(u)$ and $\varphi_{FZ}(u)$.
It is always $2nM$ in the broken ${\bf Z}_{N}$-symmetric model,
and in the Fateev-Zamolodchikov model it is $2nM-2|q|$,
which depends on the sector of
the ${\bf Z}_{N}$-charge operator ${\cal Q}$.
This difference
originates in that the ${\bf Z}_{N}$-symmetry holds only in the
Fateev-Zamolodchikov model and that it breaks
away from the criticality.
The BAE's (\ref{eq:BAE}) for the broken ${\bf Z}_{N}$-symmetric
model should coincide with those for the Fateev-Zamolodchikov
model in the trigonometric limit $\kappa\rightarrow\infty$.
We conjecture that the situation is the following.
In the solution $\{v_{1},\cdots,v_{2nM}\}$
to the BAE's (\ref{eq:BAE}) for the broken ${\bf Z}_{N}$-symmetric model,
some of them diverge to $\pm\sqrt{-1}\infty$
all in the same order in $\kappa$
when the trigonometric limit is taken.
The half of them diverge to $\sqrt{-1}\infty$, and the
other half to $-\sqrt{-1}\infty$.
The number of them are always even and between $0$ and $2n$.
Let $2q$ be this number.
Then $q$ determines the sector of the ${\bf Z}_{N}$-charge
operator in which this eigenvalue falls.
In this situation, the BAE's (\ref{eq:BAE})
and the eigenvalue $\varphi(u)$ in (\ref{eq:eigenvalue2})
surely become the BAE's (\ref{eq:AlbertiniBAE})
and $\varphi_{FZ}(u)$ in (\ref{eq:Albertinivarphi})
respectively in the trigonometric limit.

The free energy (\ref{eq:energy}) agrees with the result
of \cite{JMO86} by the inversion-trick.
The string hypothesis for the ground state in Section 5
is consistent with their result.
%
In our formulation, it is manifest that the free energy
$F(u)$ is  doubly periodic in $u$,
\begin{displaymath}
 F(u+1)=F(u+\tau/2)=F(u),
\end{displaymath}
from (\ref{eq:phiperiod}), (\ref{eq:freeenergy})
and the fact that
the ground state belongs to the sector of $r=1$.
This result of $F(u)$ also gives the ground state energy of
the one dimensional spin chain Hamiltonian ${\cal H}$
corresponding to the broken ${\bf Z}_{N}$-symmetric model,
\begin{displaymath}
  \log \Phi(u)=Id + u{\cal H} + O(u^{2}).
\end{displaymath}
The Hamiltonian ${\cal H}$ itself is modular invariant,
${\cal H}(\tau)={\cal H}(-\dfrac{1}{\tau})$.
We will report on these accounts elsewhere.

\appendix
\section{Theta Function}
\setcounter{equation}{0}
\label{ap:theta}
We summarize the necessary facts about theta functions
in this Appendix. See \cite{Mumford83}\cite{WW58} for proofs.
We define $\theta_{1}(u|\tau)$ by
\begin{displaymath}
 \theta_{1}(u|\tau)
 =
 2q^{1/4}\sin(\pi u)\prod_{n=1}^{\infty}
 (1-2q^{2n}\cos(2\pi u)+q^{4n})(1-q^{2n}),
\end{displaymath}
where $q=\exp(\sqrt{-1}\pi\tau)$.
It is an odd function in $u$ and has the quasi-periodicity
\begin{eqnarray*}
 \theta_{1}(u+1|\tau)
 &=&
 \theta_{1}(u|\tau),\\
 \theta_{1}(u+\tau|\tau)
 &=&
 -q^{-1}\exp(2\sqrt{-1}\pi u)\,\theta_{1}(u|\tau),
\end{eqnarray*}
and satisfies
\begin{equation}
 \theta_{1}(u|\tau)
 =
 \sqrt{-1}\,\Bigl(\dfrac{\sqrt{-1}}{\tau}\Bigr)^{1/2}
 \exp\Bigl(-\sqrt{-1}\pi\dfrac{u^{2}}{\tau}\Bigr)
 \theta_{1}\Bigl(\dfrac{u}{\tau}|-\dfrac{1}{\tau}\Bigr).
 \label{eq:conjugate}
\end{equation}
The other theta functions $\theta_{2}$, $\theta_{3}$ and $\theta_{4}$
are defined by
\begin{eqnarray*}
 \theta_{2}(u|\tau)
 &=&
 \theta_{1}(u+1/2|\tau),\\
 \theta_{3}(u|\tau)
 &=&
 -q^{1/4}\exp(\sqrt{-1}\pi u)\,\theta_{1}(u+1/2+\tau/2|\tau),\\
 \theta_{4}(u|\tau)
 &=&
 -\sqrt{-1}q^{1/4}\exp(\sqrt{-1}\pi u)\,\theta_{1}(u+\tau/2|\tau).
\end{eqnarray*}
We abbreviate a product of theta functions of the same argument to,
for example,
\begin{eqnarray*}
 &
 {[}\theta_{1}\theta_{2}{]}(u|\tau)
 =\theta_{1}(u|\tau)\,\theta_{2}(u|\tau),
 & \\
 &
 {[}\theta_{2}\theta_{3}\theta_{4}{]}(0|\tau)
 =\theta_{2}(0|\tau)\,\theta_{3}(0|\tau)\,\theta_{4}(0|\tau).
 &
\end{eqnarray*}
In this notation,
\begin{equation}
 \begin{array}{cc}
  {[}\theta_{1}\theta_{4}{]}(u|\tau)
   =\dfrac{\eta(2\tau)^{2}}{\eta(\tau)}\,\theta_{1}(u|\tau/2),&
  {[}\theta_{2}\theta_{3}{]}(u|\tau)
   =\dfrac{\eta(2\tau)^{2}}{\eta(\tau)}\,\theta_{2}(u|\tau/2),\\
  {[}\theta_{1}\theta_{2}{]}(u|\tau)
   =\dfrac{\eta(2\tau)^{2}}{\eta(4\tau)}\,\theta_{1}(2u|2\tau),&
  {[}\theta_{3}\theta_{4}{]}(u|\tau)
   =\dfrac{\eta(2\tau)^{2}}{\eta(4\tau)}\,\theta_{4}(2u|2\tau)
 \end{array}
 \label{eq:tautau}
\end{equation}
hold, where
\begin{displaymath}
 \eta(\tau)=q^{1/24}\prod_{n=1}^{\infty}(1-q^{n})
\end{displaymath}
is the Dedekind eta function.
The necessary addition formulae are
\begin{equation}
 \begin{array}{c}
 {[}\theta_{2}\theta_{3}{]}(u|\tau)\,
 {[}\theta_{2}\theta_{3}{]}(v|\tau) \pm
 {[}\theta_{1}\theta_{4}{]}(u|\tau)\,
 {[}\theta_{1}\theta_{4}{]}(v|\tau)
 =
 {[}\theta_{2}\theta_{3}{]}(0|\tau)\,
 \theta_{2}(u\mp v|\tau)\,\theta_{3}(u\pm v|\tau),\\
 {[}\theta_{1}\theta_{4}{]}(u|\tau)\,
 {[}\theta_{2}\theta_{3}{]}(v|\tau) \pm
 {[}\theta_{2}\theta_{3}{]}(u|\tau)\,
 {[}\theta_{1}\theta_{4}{]}(v|\tau)
 =
 {[}\theta_{2}\theta_{3}{]}(0|\tau)\,
 \theta_{1}(u\pm v|\tau)\,\theta_{4}(u\mp v|\tau).
 \end{array}
\end{equation}
In Section 5, we use the Fourier expansions,
\begin{eqnarray}
 &
 \dfrac{1}{\sqrt{-1}}\log
 \dfrac{\theta_{1}(w+v|\tau)}{\theta_{1}(w-v|\tau)}
 =
 -2\pi \{v\}
 +2\sum_{k=1}^{\infty}
 \dfrac{\sin(2\pi kv)\,\sin(2\pi k(w-\tau/2))}
       {k\, \sin(\pi k\tau)},
 & \label{eq:logexpan}\\
 &
 \dfrac{\theta_{3}(u|\tau)}{\theta_{4}(u|\tau)}
 =
 \dfrac{1}{ {[}\theta_{3}\theta_{4}{]}(0|\tau)}
 \sum_{k=-\infty}^{\infty}
 \dfrac{\exp(\sqrt{-1}\pi ku)}{\cos(\pi k\tau)},
 & \label{eq:dnexpan}
\end{eqnarray}
which are valid for $0 < \mbox{Im}(w) <\tau$ and $v$ real.
We are denoting the fractional part of $x$ by $\{x\}$.
The expansion (\ref{eq:dnexpan}) is  essentially the same
as that of the Jacobi
elliptic function $\mbox{dn}(u,k)$
\begin{displaymath}
 \mbox{dn}(u,k)
 =
 \dfrac{\pi}{2K}\sum_{l=-\infty}^{\infty}
 \dfrac{\exp(\sqrt{-1}\pi lu/K)}{\cos(\pi l\tau)},
 \quad K=\dfrac{\pi}{2}\,\theta_{3}(0|\tau)^{2}.
\end{displaymath}
The next lemma is fundamental.

\newtheorem{lemma}{Lemma}
\begin{lemma}
\label{lemma}
 Let $f(u)$ be a meromorphic function which is not identically zero
 and has the quasi-periodicity property
 \begin{eqnarray*}
  &
  f(u+1)=\exp\Bigl(-2\sqrt{-1}\pi B\Bigr)\, f(u),
  & \\
  &
  f(u+\tau)=\exp\Bigl(-2\sqrt{-1}\pi(A_{1}+A_{2}u)\Bigr)\,f(u).
  &
 \end{eqnarray*}
 Denoting the zeros and poles of f(u) by
 $u_{1},u_{2},\cdots,u_{n}$ and
 $v_{1},v_{2},\cdots,v_{n}$ respectively, then we have
 \begin{displaymath}
  n-m=A_{2},
  \quad
  \dsum{j=1}{n}u_{j}-\dsum{j=1}{m}v_{j}
  \equiv \dfrac{1}{2}A_{2}-A_{1}+B\tau
  \quad \bmod ({\bf Z}\oplus \tau{\bf Z}),
 \end{displaymath}
and
 \begin{displaymath}
  f(u)=C \,\exp\Bigl(\sqrt{-1}\pi(A_{2}-2B)u\Bigr)
       \dfrac{\displaystyle{\prod_{j=1}^{n}}
              \theta_{1}(u-u_{j}|\tau)}
             {\displaystyle{\prod_{j=1}^{m}}
              \theta_{1}(u-v_{j}|\tau)}
 \end{displaymath}
with $C$ independent of $u$.
\end{lemma}

\section{Boltzmann Weights}
\setcounter{equation}{0}
\label{ap:Boltz}
In this Appendix, we list some formulae about
the Boltzmann weights.
Solving the recursion relations (\ref{eq:recW}) and (\ref{eq:recWb})
under the normalization of
$W(0,0|u)$ $=$ $\overline{W}(0,0|u)$ $=$ $1$,
we have, for $a,b=0,1,\cdots,n$,
\begin{displaymath}
 W(2a,2b|u) = W(N-2a,N-2b|u)
            = T\mbox{}^{(+)}_{2}(|a-b||u)\,T\mbox{}^{(+)}_{3}(a+b|u),
\end{displaymath}
\begin{displaymath}
 W(2a,N-2b|u) = W(N-2a,2b|u)
            = T\mbox{}^{(+)}_{2}(a+b|u)\,T\mbox{}^{(+)}_{3}(|a-b||u),
\end{displaymath}
\begin{displaymath}
 \overline{W}(2a,2b|u) = \overline{W}(N-2a,N-2b|u)
     = G_{2a}G_{2b}T\mbox{}^{(-)}_{2}(|a-b||u)\,T\mbox{}^{(-)}_{3}(a+b|u),
\end{displaymath}
\begin{displaymath}
 \overline{W}(2a,N-2b|u) = \overline{W}(N-2a,2b|u)
     = G_{2a}G_{2b}T\mbox{}^{(-)}_{2}(a+b|u)\,T\mbox{}^{(-)}_{3}(|a-b||u).
\end{displaymath}
Noting that $T^{(\sigma)}_{k}(a|u)$ in (\ref{eq:T}) satisfies
\begin{eqnarray}
 &
 T^{(\sigma)}_{k}(0|u)=T^{(\sigma)}_{k}(N|u),
 & \label{eq:t1} \\
 &
 T^{(\sigma)}_{k}(N-a|u)=T^{(\sigma)}_{k}(a|u),
 & \label{eq:t2}
\end{eqnarray}
we can extend the domain of the first argument
of $T^{(\sigma)}_{k}$ to all integers by periodicity.
With this convention we can rewrite the above
expression of the Boltzmann weights simply as
\begin{eqnarray*}
 W(2a,2b|u)
 &=&
 T^{(+)}_{2}(a-b|u)\,T^{(+)}_{3}(a+b|u), \\
 \overline{W}(2a,2b|u)
 &=&
 G_{2a}\,G_{2b}\,T^{(-)}_{2}(a-b|u)\,T^{(-)}_{3}(a+b|u).
\end{eqnarray*}
We have in particular at $u$$=$$0$ and $\lambda$,
\begin{equation}
 W(a,b|0)=G_{a}^{-1}G_{b}^{-1}\overline{W}(a,b|\lambda)=1,
 \quad G_{a}\,G_{b}W(a,b|\lambda)=\overline{W}(a,b|0)=\delta_{ab}.
 \label{eq:WWbzero}
\end{equation}
When we write $u+a\eta$ as $U_{a}$,
the next identity for $T^{(\sigma)}_{k}$
\begin{displaymath}
 \dfrac{T^{(\sigma_{1})}_{k}(n-a|u)}
       {T^{(\sigma_{1})}_{k}(a|u+\sigma_{2}\eta)}
 =
 \Biggl(\dfrac{\theta_{k}(U_{(1+\sigma_{1})/2})}
              {\theta_{k}(U_{-2\sigma_{2}+(1+\sigma_{1})/2})}
 \Biggr)^{\sigma_{1}\sigma_{2}}\,
 T^{(\sigma_{1})}_{k}(n|u)
\end{displaymath}
gives
\begin{equation}
 \begin{array}{l}
 \dfrac{W(a,b-1|u)}
       {W(a,b|u+\sigma\eta)}
 =
 \biggl(\theta_{2}\Bigl(U_{\sigma(-a+b)}\Bigr)
        \theta_{3}\Bigl(U_{\sigma(a+b)}\Bigr)
 \biggr)^{-\sigma}
 {[}T^{(+)}_{2}T^{(+)}_{3}{]}(n|u),
 \\
 \dfrac{\overline{W}(a,b-1|u)}
       {\overline{W}(a,b|u+\sigma\eta)}
 =
 \dfrac{G_{b-1}}{G_{b}}\,
 \biggl(\theta_{1}\Bigl(U_{\sigma(-a+b)+1}\Bigr)
       \theta_{4}\Bigl(U_{\sigma(a+b)+1}\Bigr)
 \biggr)^{\sigma}
 {[}T^{(-)}_{2}T^{(-)}_{3}{]}(n|u).
 \end{array}
 \label{eq:identity1}
\end{equation}
$T^{(\sigma)}_{k}$ has the quasi-periodicity
\begin{eqnarray*}
 &
 T^{(\sigma)}_{k}(a|u+1)
 =
 \exp(4\sqrt{-1}\pi\sigma a^{2}\eta)\,
 T^{(\sigma)}_{k}(a|u+\tau)
 =
 T^{(\sigma)}_{k}(a|u),
 & \label{eq:Tperiod1} \\
 &
 T^{(\sigma)}_{k}(a|u+\tau/2)
 =
 \exp(-2\sqrt{-1}\pi\sigma a^{2}\eta)\,
 T^{(\sigma)}_{k'}(a|u),
 \,k+k'\equiv 0 \bmod 5.
 & \label{eq:Tperiod2}
\end{eqnarray*}
Hence we have
\begin{equation}
 W(a,b,|u+1)=W(a,b|u),\quad
 \overline{W}(a,b,|u+1)=\overline{W}(a,b|u),
 \label{eq:Wperiod1}
\end{equation}
and
\begin{displaymath}
 \begin{array}{l}
 W(a,b|u+\tau/2)
 =
 \exp(-4\sqrt{-1}\pi(a^{2}+b^{2})\eta)\,
 W(a,N-b|u),\\
 \overline{W}(a,b|u+\tau/2)
 =
 \exp(4\sqrt{-1}\pi(a^{2}+b^{2})\eta)\,
 \overline{W}(a,N-b|u).
 \end{array}
 \label{eq:Wperiod2}
\end{displaymath}
We note
\begin{eqnarray}
 \lefteqn{W(a,b|u)\,\overline{W}(b,c|u)} \nonumber \\
 &=
 \exp(-4\sqrt{-1}\pi(a^{2}-c^{2})\eta)\,
 W(a,N-b|u+\tau/2)\,\overline{W}(N-b,c|u+\tau/2).&
 \label{eq:WWbperiod}
\end{eqnarray}

\section{Commutation Relation between ${\cal L}$ and $\Phi$}
\setcounter{equation}{0}
\label{ap:ComRel}
In this appendix, we give a proof of (\ref{eq:LPHI1}).
A graphical representation of this proof for the case of $M=2$
is illustrated in $Fig.\ 11$.
\begin{eqnarray*}
 & &
 \Bigl(\Phi(u,v,w)\,{\cal L}(u,v,w')\Bigr)\mbox{}
 ^{c_{0}\cdots c_{M-1}}_{a_{0}\cdots a_{M-1}} \\
 &=&
 \dsum{b_{0}\cdots b_{M-1}}{}
 \Phi(u,v,w)^{c_{0}\cdots c_{M-1}}_{b_{0}\cdots b_{M-1}}
 {\cal L}(u,v,w')^{b_{0}\cdots b_{M-1}}_{a_{0}\cdots a_{M-1}}\\
 &=&
 \dsum{{b_{0}\cdots b_{M-1} \atop i_{0}\cdots i_{M-1}}}{}
 \Bigl(\dprod{j=0}{M-1}
        W(b_{j-1},c_{j}|u-w)\,
        \overline{W}(c_{j},b_{j}|v-w)\,
        K\mbox{}^{\ \ \ \ b_{j}}_{i_{j+1}a_{j}}(u-w')\,
        K\mbox{}^{i_{j}b_{j}}_{\ \  a_{j}}(v-w')
 \Bigr).
\end{eqnarray*}
By the unitarity relation (\ref{eq:unitarity2}), inserting
\begin{displaymath}
 1=\dsum{i'}{}\delta_{i_{0}i'}
 =
 \dfrac
 {\dsum{i'a'}{}G_{a'}\,
  K\mbox{}^{\ c_{0}}_{i'a'}(w-w'+\lambda)
  K\mbox{}^{i_{0}c_{0}}_{\ \ a'}(w-w')}
 {{[}\theta_{2}\theta_{3}\theta_{4}{]}(0)\,
  G_{c_{0}}\theta_{2}(2w-2w')},
\end{displaymath}
we have
\begin{eqnarray*}
 &=&
 \Bigl({[}\theta_{2}\theta_{3}\theta_{4}{]}(0)\,
        G_{c_{0}}\,\theta_{2}(2w-2w')\Bigr)^{-1}
 \dsum{{b_{0}\cdots b_{M-1} \atop i_{0}\cdots i_{M-1}}}{}
 \dsum{i'a'}{}
 G_{a'}\,K\mbox{}^{\ c_{0}}_{i'a'}(w-w'+\lambda)\\
 & &
 \times\Bigl(\dprod{j=0}{M-1}
        \overline{W}(c_{j},b_{j}|v-w)\,
        W(b_{j-1},c_{j}|u-w)
 \Bigr)
 K\mbox{}^{i_{0}c_{0}}_{\ \ a'}(w-w')
 K\mbox{}^{\ \ b_{M-1}}_{i_{0}a_{M-1}}(u-w')\\
 & &
 \times\Bigl( \dprod{j=1}{M-1}
         K\mbox{}^{i_{j}b_{j}}_{\ \  a_{j}}(v-w')\,
         K\mbox{}^{\ \ b_{j-1}}_{i_{j}a_{j-1}}(u-w')\Bigr)
 K\mbox{}^{i'b_{0}}_{\ a_{0}}(v-w').
\end{eqnarray*}
Successive use of (\ref{eq:defW}) and (\ref{eq:defWb})
yields
\begin{eqnarray*}
 &=&
 \Bigl({[}\theta_{2}\theta_{3}\theta_{4}{]}(0)\,
        G_{c_{0}}\,\theta_{2}(2w-2w')\Bigr)^{-1}\\
 & &
 \times\dsum{{b_{0}\cdots b_{M-1} \atop i_{0}\cdots i_{M-1}}}{}
 \dsum{i'a'}{}
 G_{a'}\,K\mbox{}^{\ c_{0}}_{i'a'}(w-w'+\lambda)\,
 K\mbox{}^{i'c_{0}}_{\ \, b_{0}}(w-w')\\
 & &
 \times\Bigl( \dprod{j=0}{M-2}
         K\mbox{}^{\ \ c_{j}}_{i_{j}b_{j}}(v-w')\,
         K\mbox{}^{i_{j}c_{j+1}}_{\ \  b_{j+1}}(u-w')
 \Bigr)
 K\mbox{}^{\ \ \ \ \ c_{M-1}}_{i_{M-1}b_{M-1}}(v-w')\,
 K\mbox{}^{i_{M-1}c_{0}}_{\ \ \ \ \ a'}(u-w')\\
 & &
 \times\Bigl(\dprod{j=0}{M-2}
        \overline{W}(b_{j},a_{j}|v-w)\,
        W(a_{j},b_{j+1}|u-w)
 \Bigr)\\
 & &
 \times\overline{W}(b_{M-1},a_{M-1}|v-w)\,
 W(a_{M-1},a'|u-w).
\end{eqnarray*}
By the unitarity relation (\ref{eq:unitarity1}), we have
\begin{displaymath}
 \dfrac
 {\dsum{i'}{}G_{a_{0}}\,
  K\mbox{}^{\ c_{0}}_{i'a'}(w-w'+\lambda)
  K\mbox{}^{i'\ c_{0}}_{\ b_{M-1}}(w-w')}
 {{[}\theta_{2}\theta_{3}\theta_{4}{]}(0)\,
  G_{c_{0}}\theta_{2}(2w-2w')}
 =
 \delta_{a'b_{M-1}},
\end{displaymath}
then the above formula reduces
\begin{eqnarray*}
 &=&
 \dsum{{b_{0}\cdots b_{M-1} \atop i_{0}\cdots i_{M-1}}}{}
 \Bigl(\dprod{j=0}{M-1}
        K\mbox{}^{\ \ c_{j}}_{i_{j}b_{j}}(v-w')\,
        K\mbox{}^{i_{j}c_{j+1}}_{\ \  b_{j+1}}(u-w')\,
        \overline{W}(b_{j},a_{j}|v-w)\,
        W(a_{j},b_{j+1}|u-w)\,
 \Bigr)\\
 &=&
 \dsum{b_{0}\cdots b_{M-1}}{}
 {\cal L}(u,v,w')^{c_{0}\cdots c_{M-1}}_{b_{0}\cdots b_{M-1}}
 \Phi(u,v,w)^{b_{0}\cdots b_{M-1}}_{a_{0}\cdots a_{M-1}}\\
 &=&
 \Bigl({\cal L}(u,v,w')\,\Phi(u,v,w)\Bigr)\mbox{}
 ^{c_{0}\cdots c_{M-1}}_{a_{0}\cdots a_{M-1}}.
\end{eqnarray*}
Now we obtain
\begin{displaymath}
 \Phi(u,v,w)\,{\cal L}(u,v,w')
 =
 {\cal L}(u,v,w')\,\Phi(u,v,w).
\end{displaymath}

\section*{Acknowledgment}
The author would like to express his gratitude to Prof. M. Kashiwara,
Prof. T. Miwa and Prof. M. Jimbo for their encouragement.
Thanks are also due to Dr Y-.H. Quano for his careful reading of
the manuscript.

\pagebreak[4]

\pagebreak[4]
\Large{
\begin{center}
 Figure caption
\end{center}
\begin{tabular}{cl}
Fig1 & Graphical representation of \\
     &            $W(a,b|u,v)$ and $\overline{W}(a,b|u,v)$      \\
Fig2 & Graphical representation of the STR      \\
Fig3 & Graphical representation of $\Phi(u,v,w)$   \\
Fig4 & Graphical representation of ($1.10$)\\
Fig5 & Graphical representation of the YBE\\
Fig6 & Graphical representations of \\
     &            $K_{ia}^{\ b}(u-w)$ and $K_{\ a}^{jb}(v-w)$\\
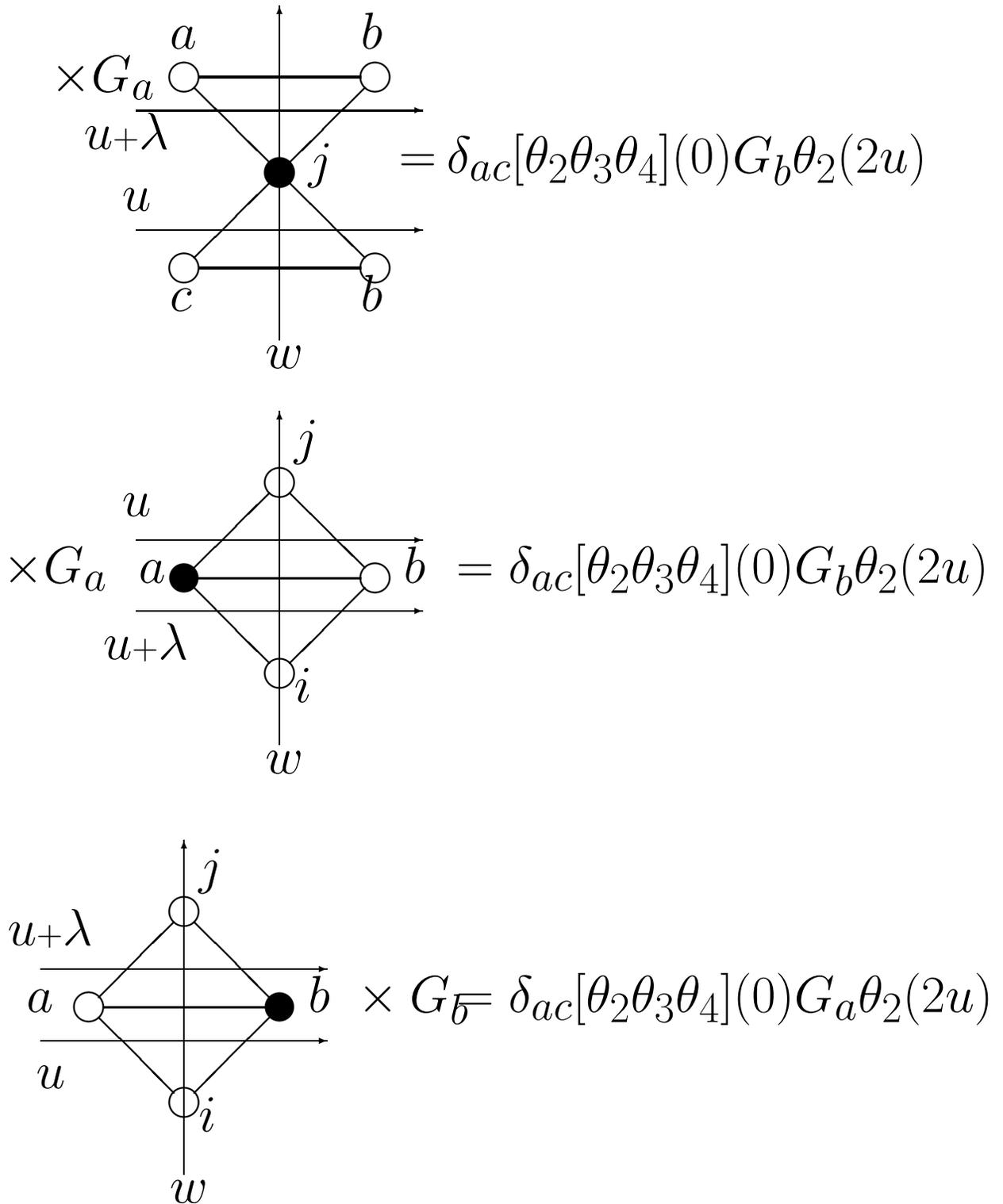
Fig7 & Graphical representations of \\
     & unitary relations between $K$'s\\
Fig8 & Graphical representation of ($2.9$)\\
Fig9 & Graphical representation of ($2.10$)\\
Fig10 & Graphical representations of ${\cal L}(u,v,w)$ \\
Fig11 & Graphical representation of a proof of \\
      &              ${\cal L}(u,v,w')$ $\Phi(u,v,w)$
                  $=$ $\Phi(u,v,w)$ ${\cal L}(u,v,w')$
\end{tabular}
}

\pagebreak[4]



\setlength{\unitlength}{1mm}
\begin{picture}(145,230)
      \put(10,170){
      \begin{picture}(20,100)
      \put(-10,0){\Huge $W(a,b|u,v)$}
      \put(45,0){\Huge $=$}
      \thicklines
      \put(95,-17){\line(0,1){34}}
      \put(95,-20){\circle{6}}\put(92,-29){\Huge $a$}
      \put(95,20){\circle{6}} \put(93,24){\Huge $b$}
      \thinlines
      \put(80,-15){\vector(1,1){30}}  \put(75,-18){\Huge $v$}
      \put(80,15){\vector(1,-1){30}} \put(75,16){\Huge $u$}
      \end{picture}
      }
      \put(10,90){
      \begin{picture}(20,100)
      \put(-10,-2){\Huge $\overline{W}(a,b|u,v)$}
      \put(45,0){\Huge $=$}
      \thicklines
      \put(78,1){\line(1,0){34}}
      \put(78,-1){\line(1,0){34}}
      \put(75,0){\circle{6}} \put(66,-1){\Huge $a$}
      \put(115,0){\circle{6}}\put(121,-1){\Huge $b$}
      \thinlines
      \put(80,-15){\vector(1,1){30}}  \put(75,-18){\Huge $v$}
      \put(80,15){\vector(1,-1){30}}  \put(75,16){\Huge $u$}
      \end{picture}
      }
      \put(0,20){\Huge {\rm Fig}.1 \ Graphical representations of}
      \put(40,11){\Huge $W(a,b|u,v)$ and $\overline{W}(a,b|u,v)$}
\end{picture}


\setlength{\unitlength}{1mm}

\begin{picture}(145,230)
  \thicklines
  \put(20,170){\circle{6}}
  \put(20,130){\circle{6}}
  \put(50,150){\circle{6}}
  \put(18.5,132.5){\line(0,1){35}}
  \put(21.5,132.5){\line(0,1){35}}
  \put(22.5,169){\line(3,-2){25.5}}
  \put(22.5,131){\line(3,2){25.5}}
  \thinlines
  \put(10,144){\vector(3,2){40}}
  \put(50,129){\vector(-3,2){40}}
  \put(35,120){\vector(0,1){55}}
  \put(18,175){\Huge $c$}
  \put(18,122){\Huge $a$}
  \put(55,147){\Huge $b$}
  \put(8,137){\Huge $u$}
  \put(32,115){\Huge $v$}
  \put(52,124){\Huge $w$}
  \put(0,147){\Huge $\rho$\large$\ \times$}

  \put(70,147){\Huge $=$}

  \put(90,171){\circle{6}}
  \put(90,129){\circle{6}}
  \put(104,150){\circle*{6}}
  \put(92.5,169.7){\line(2,-3){11.5}}
  \put(91,168.4){\line(2,-3){11.5}}
  \put(92.5,130.3){\line(2,3){11.5}}
  \put(91,131.6){\line(2,3){11.5}}
  \put(130,150){\circle{6}}
  \put(106,150){\line(1,0){21}}
  \thinlines
  \put(127,145){\vector(-2,1){40}}
  \put(87,135){\vector(2,1){40}}
  \put(98,127){\vector(0,1){46}}
  \put(88,176){\Huge $c$}
  \put(88,121){\Huge $a$}
  \put(135,147){\Huge $b$}
  \put(80,132){\Huge $u$}
  \put(97,120){\Huge $v$}
  \put(130,138){\Huge $w$}
  \put(-10,70){\Huge Fig.\ 2\ Graphical representation of the STR}

\end{picture}



 \setlength{\unitlength}{1mm}

 \begin{picture}(145,230)
       \multiput(90,50)(0,40){2}
	  {
	  \begin{picture}(40,40)
	    \thicklines
	    \put(0,20){\circle{6}}
	    \put(2.12,17.88){\line(1,-1){15.76}}
	    \put(20,0){\circle{6}}
	    \put(1.08,23.0){\line(1,1){15.95}}
	    \put(3.0,21.12){\line(1,1){15.86}}
	    \put(20,40){\circle{6}}
	    \thinlines
	    \put(-20,10){\vector(1,0){60}}
	    \put(-26,8){\Huge $u$}
	    \put(-20,30){\vector(1,0){60}}
	    \put(-26,30){\Huge $v$}
	  \end{picture}
	   }
       \put(90,170)
	  {
	  \begin{picture}(40,40)
	    \thicklines
	    \put(0,20){\circle{6}}
	    \put(2.12,17.88){\line(1,-1){15.76}}
	    \put(20,0){\circle{6}}
	    \put(1.08,23.0){\line(1,1){15.95}}
	    \put(3.0,21.12){\line(1,1){15.86}}
	    \put(20,40){\circle{6}}
	    \thinlines
	    \put(-20,10){\vector(1,0){60}}
	    \put(-26,8){\Huge $u$}
	    \put(-20,30){\vector(1,0){60}}
	    \put(-26,30){\Huge $v$}
	  \end{picture}
	   }
       \thinlines
       \put(103,30){\vector(0,1){200}}
       \put(98,24){\Huge $w$}
       \put(73,70){\Huge $a$\Large$\mbox{}_{M-1}$}
       \put(73,110){\Huge $a$\Large$\mbox{}_{M-2}$}
       \put(78,190){\Huge $a_{0}$}
       \put(117,48){\Huge $b_{0}$}
       \put(117,88){\Huge $b$\Large$\mbox{}_{M-1}$}
       \put(117,128){\Huge $b$\Large$\mbox{}_{M-2}$}
       \put(117,168){\Huge $b_{1}$}
       \put(117,208){\Huge $b_{0}$}
       \multiput(80,117)(0,4){18}{\circle*{0.8}}
       \multiput(98,125)(0,4){14}{\circle*{0.8}}
       \multiput(120,137)(0,4){8}{\circle*{0.8}}
       \put(-15,130){\Huge $\Phi(u,v,w)_{a_{0}a_{1}\cdots a_{M-1}}
		     ^{b_{0}b_{1}\cdots b_{M-1}} =$}
       \put(-15,0){\Huge {\rm Fig}.\ 3 \ Graphical representation
		  of $\Phi(u,v,w)$}
 \end{picture}



\setlength{\unitlength}{1mm}

\begin{picture}(145,230)
 \put(30,-120){
 \put(40,180){
      \begin{picture}(40,40)
      \put(0,0){\vector(1,1){45}}
      \put(40,0){\vector(-1,1){45}}
      \put(0,40){\circle{6}}
      \put(-5,31){\Huge $j_{3}$}
      \put(40,40){\circle{6}}
      \put(42,32){\Huge $i_{3}$}
      \end{picture}
            }
 \put(60,140){
      \begin{picture}(40,40)
      \thicklines
      \put(3,20){\line(1,0){34}}
      \put(0,20){\circle*{8}}
      \put(2.12,22.12){\line(1,1){15.73}}
      \put(40,20){\circle{6}} \put(46,18){\Huge $c$}
      \put(2.12,17.88){\line(1,-1){15.73}}
      \put(20,0){\circle{6}} \put(24,-9){\Huge $j_{1}$}
      \put(37.88,22.12){\line(-1,1){15.73}}
      \put(20,40){\circle*{8}} \put(24,46){\Huge $j_{2}$}
      \put(37.88,17.88){\line(-1,-1){15.73}}
      \thinlines
      \put(-7,32){\vector(1,0){55}}
      \put(-7,8){\vector(1,0){55}}
      \put(20,-10){\vector(0,1){50}} \put(18,-16){\Huge $w$}
      \end{picture}
              }
 \put(20,140){
      \begin{picture}(40,40)
      \thicklines
      \put(3,20){\line(1,0){34}}
      \put(0,20){\circle{6}} \put(-8,20){\Huge $a$}
      \put(2.12,22.12){\line(1,1){15.73}}
      \put(-2,24){\Huge $b$}
      \put(2.12,17.88){\line(1,-1){15.73}}
      \put(20,0){\circle{6}} \put(24,-9){\Huge $i_{1}$}
      \put(37.88,22.12){\line(-1,1){15.73}}
      \put(20,40){\circle*{8}} \put(13,46){\Huge $i_{2}$}
      \put(37.88,17.88){\line(-1,-1){15.73}}
      \thinlines
      \put(-7,32){\vector(1,0){55}}
      \put(-13,30){\Huge $u$}
      \put(-7,8){\vector(1,0){55}}
      \put(-13,6){\Huge $u$}
      \put(20,-10){\vector(0,1){50}} \put(18,-16){\Huge $v$}
      \end{picture}
              }
           }
 \put(15,58){\Huge $=$}
 \put(-20,100){
 \put(60,20){
      \begin{picture}(40,40)
      \put(-5,-5){\vector(1,1){45}}
      \put(45,-5){\vector(-1,1){45}}
      \put(0,0){\circle{6}}
      \put(-4,7){\Huge $i_{1}$}
      \put(40,0){\circle{6}}
      \put(40,7){\Huge $j_{1}$}
      \end{picture}
            }
 \put(40,60){
      \begin{picture}(40,40)
      \thicklines
      \put(3,20){\line(1,0){34}}
      \put(0,20){\circle{6}} \put(-8,18){\Huge $a$}
      \put(2.12,22.12){\line(1,1){15.73}}
      \put(38,24){\Huge $b$}
      \put(2.12,17.88){\line(1,-1){15.73}}
      \put(20,0){\circle*{8}}  \put(18,-9){\Huge $j_{2}$}
      \put(37.88,22.12){\line(-1,1){15.73}}
      \put(20,40){\circle{6}} \put(24,46){\Huge $j_{3}$}
      \put(37.88,17.88){\line(-1,-1){15.73}}
      \thinlines
      \put(-7,32){\vector(1,0){55}}
      \put(-13,30){\Huge $u$}
      \put(-7,8){\vector(1,0){55}}
      \put(-13,6){\Huge $u$}
      \put(20,0){\vector(0,1){50}} \put(18,54){\Huge $w$}
      \end{picture}
              }
 \put(80,60){
      \begin{picture}(40,40)
      \thicklines
      \put(3,20){\line(1,0){34}}
      \put(0,20){\circle*{8}}
      \put(2.12,22.12){\line(1,1){15.73}}
      \put(40,20){\circle{6}} \put(46,19){\Huge $c$}
      \put(2.12,17.88){\line(1,-1){15.73}}
      \put(20,0){\circle*{8}}  \put(24,-9){\Huge $i_{2}$}
      \put(37.88,22.12){\line(-1,1){15.73}}
      \put(20,40){\circle{6}} \put(24,46){\Huge $i_{3}$}
      \put(37.88,17.88){\line(-1,-1){15.73}}
      \thinlines
      \put(-7,32){\vector(1,0){55}}
      \put(-7,8){\vector(1,0){55}}
      \put(20,0){\vector(0,1){50}} \put(18,54){\Huge $v$}
      \end{picture}
              }
          }

\put(-5,-10){\Huge {\rm Fig}.\ 4 \ Graphical representation of (\ref{eq:LLR})}

\end{picture}


\setlength{\unitlength}{1mm}

\begin{picture}(145,230)
  \put(20,135){\vector(0,1){70}}
  \put(0,150){\vector(2,1){60}}
  \put(60,155){\vector(-2,1){60}}
  \put(20,148){\circle{4}}
  \put(20,168){\circle*{4}}
  \put(20,188){\circle{4}}
  \put(27,172){\circle*{4}}
  \put(7,182){\circle{4}}
  \put(47,162){\circle{4}}
  \put(27,163){\circle*{4}}
  \put(7,153){\circle{4}}
  \put(47,173){\circle{4}}

  \put(-3,145){\Huge $u$}
  \put(18,130){\Huge $v$}
  \put(60,150){\Huge $w$}
  \put(2,158){\Huge $i_{1}$}
  \put(27,155){\Huge $i_{2}$}
  \put(43,180){\Huge $i_{3}$}
  \put(25,143){\Huge $j_{1}$}
  \put(10,165){\Huge $j_{2}$}
  \put(25,190){\Huge $j_{3}$}
  \put(45,152){\Huge $k_{1}$}
  \put(27,177){\Huge $k_{2}$}
  \put(2,174){\Huge $k_{3}$}

  \put(72,165){\Huge $=$}

  \put(130,135){\vector(0,1){70}}
  \put(150,150){\vector(-2,1){60}}
  \put(90,155){\vector(2,1){60}}
  \put(130,148){\circle{4}}
  \put(130,168){\circle*{4}}
  \put(130,188){\circle{4}}
  \put(123,172){\circle*{4}}
  \put(143,182){\circle{4}}
  \put(103,162){\circle{4}}
  \put(123,163){\circle*{4}}
  \put(143,153){\circle{4}}
  \put(103,173){\circle{4}}

  \put(153,145){\Huge $w$}
  \put(132,130){\Huge $v$}
  \put(90,150){\Huge $u$}
  \put(143,158){\Huge $k_{1}$}
  \put(118,155){\Huge $k_{2}$}
  \put(102,180){\Huge $k_{3}$}
  \put(120,143){\Huge $j_{1}$}
  \put(135,167){\Huge $j_{2}$}
  \put(120,190){\Huge $j_{3}$}
  \put(102,152){\Huge $i_{1}$}
  \put(118,177){\Huge $i_{2}$}
  \put(145,174){\Huge $i_{3}$}

  \put(20,68){\Huge $R\mbox{}_{ij}^{kl}(u-v)$}
  \put(72,68){\Huge $=$}
  \put(100,70){\vector(1,0){40}}
  \put(120,50){\vector(0,1){40}}
  \put(110,70){\circle{4}}
  \put(130,70){\circle{4}}
  \put(120,60){\circle{4}}
  \put(120,80){\circle{4}}
  \put(93,68){\Huge $u$}
  \put(118,42){\Huge $v$}
  \put(107,61){\Huge $i$}
  \put(113,55){\Huge $j$}
  \put(132,75){\Huge $k$}
  \put(125,82){\Huge $l$}

  \put(-10,10){\Huge Fig.\ 5\ Graphical Representation of the YBE}

\end{picture}



\setlength{\unitlength}{1mm}
\begin{picture}(145,230)
      \put(10,170){
      \begin{picture}(20,100)
      \put(0,0){\Huge $K\mbox{}_{ia}^{\ b}(u-w)$}
      \put(45,0){\Huge $=$}
      \thicklines
      \put(78,10){\line(1,0){34}}
      \put(75,10){\circle{6}} \put(72,16){\Huge $a$}
      \put(77.12,7.88){\line(1,-1){15.76}}
      \put(95,-10){\circle{6}}\put(98,-16){\Huge $i$}
      \put(112.88,7.88){\line(-1,-1){15.76}}
      \put(115,10){\circle{6}}\put(112,16){\Huge $b$}
      \thinlines
      \put(65,3){\vector(1,0){60}} \put(59,1){\Huge $u$}
      \put(95,-25){\vector(0,1){50}}\put(92,-31){\Huge $w$}
      \end{picture}
      }
      \put(10,90){
      \begin{picture}(20,100)
      \put(0,0){\Huge $K\mbox{}_{\ a}^{jb}(v-w)$}
      \put(45,0){\Huge $=$}
      \thicklines
      \put(78,-10){\line(1,0){34}}
      \put(75,-10){\circle{6}} \put(72,-19){\Huge $a$}
      \put(77.12,-7.88){\line(1,1){15.76}}
      \put(95,10){\circle{6}}\put(98,16){\Huge $j$}
      \put(97.12,7.88){\line(1,-1){15.76}}
      \put(115,-10){\circle{6}}\put(112,-19){\Huge $b$}
      \thinlines
      \put(65,-2){\vector(1,0){60}} \put(59,-3){\Huge $v$}
      \put(95,-25){\vector(0,1){50}}\put(92,-31){\Huge $w$}
      \end{picture}
      }
      \put(10,20){\Huge {\rm Fig}.\ 6\  Graphical representations of}
      \put(40,11){\Huge $K_{ia}^{\ b}(u-w)$ and $K_{\ a}^{jb}(v-w)$}
\end{picture}



\setlength{\unitlength}{0.8mm}

\begin{picture}(145,230)
      \put(-40,200){
      \begin{picture}(20,100)
      \thicklines
      \put(78,10){\line(1,0){34}}
      \put(75,10){\circle{6}} \put(72,16){\Huge $a$}
      \put(77.12,7.88){\line(1,-1){15.76}}\put(47,7){\Huge $\times G_{a}$}
      \put(95,-10){\circle*{6}}
      \put(112.88,7.88){\line(-1,-1){15.76}}
      \put(115,10){\circle{6}}\put(112,16){\Huge $b$}
      \thinlines
      \put(65,3){\vector(1,0){60}}
      \put(54,-5){\Huge$u$\Large$+$\Huge$\lambda$}
      \end{picture}
      }
      \put(-40,180){
      \begin{picture}(20,100)
      \thicklines
      \put(78,-10){\line(1,0){34}}
      \put(75,-10){\circle{6}} \put(72,-19){\Huge $c$}
      \put(77.12,-7.88){\line(1,1){15.76}}
      \put(95,10){\circle{6}}\put(101,9){\Huge $j$}
      \put(97.12,7.88){\line(1,-1){15.76}}
      \put(115,-10){\circle{6}}\put(112,-19){\Huge $b$}
      \thinlines
      \put(65,-2){\vector(1,0){60}} \put(62,2){\Huge $u$}
      \put(95,-25){\vector(0,1){70}}\put(92,-31){\Huge $w$}
      \put(120,10){\Huge $=$}
      \put(130,10){\huge $\delta_{ac}{[}\theta_{2}\theta_{3}\theta_{4}{]}(0)
                   G_{b}\theta_{2}(2u)$}
      \end{picture}
      }

      \put(-40,95){
      \begin{picture}(20,100)
      \thicklines
      \put(78,10){\line(1,0){34}}
      \put(75,10){\circle*{6}} \put(37,9){\Huge $\times G_{a}\ \ a$}
      \put(77.12,7.88){\line(1,-1){15.76}}
      \put(95,-10){\circle{6}}\put(98,-16){\Huge $i$}
      \put(112.88,7.88){\line(-1,-1){15.76}}
      \put(115,10){\circle{6}}\put(121,9){\Huge $b$}
      \thinlines
      \put(65,3){\vector(1,0){60}}
      \put(58,-7){\Huge$u$\Large$+$\Huge$\lambda$}
      \put(95,-25){\vector(0,1){70}}\put(92,-31){\Huge $w$}
      \end{picture}
      }
      \put(-40,115){
      \begin{picture}(20,100)
      \thicklines
      \put(78,-10){\line(1,0){34}}
      \put(77.12,-7.88){\line(1,1){15.76}}
      \put(95,10){\circle{6}}\put(98,16){\Huge $j$}
      \put(97.12,7.88){\line(1,-1){15.76}}
      \thinlines
      \put(65,-2){\vector(1,0){60}} \put(62,3){\Huge $u$}
      \put(132,-11){\huge $=\delta_{ac}{[}\theta_{2}\theta_{3}\theta_{4}{]}(0)
                   G_{b}\theta_{2}(2u)$}
      \end{picture}
      }

      \put(-60,5){
      \begin{picture}(20,100)
      \thicklines
      \put(78,10){\line(1,0){34}}
      \put(75,10){\circle{6}} \put(62,9){\Huge $a$}
      \put(77.12,7.88){\line(1,-1){15.76}}
      \put(95,-10){\circle{6}}\put(98,-16){\Huge $i$}
      \put(112.88,7.88){\line(-1,-1){15.76}}
      \put(115,10){\circle*{6}}\put(121,9){\Huge $b\ \times G_{b}$}
      \thinlines
      \put(65,3){\vector(1,0){60}}
      \put(64,-7){\Huge$u$}
      \put(95,-25){\vector(0,1){70}}\put(92,-31){\Huge $w$}
      \end{picture}
      }
      \put(-60,25){
      \begin{picture}(20,100)
      \thicklines
      \put(78,-10){\line(1,0){34}}
      \put(77.12,-7.88){\line(1,1){15.76}}
      \put(95,10){\circle{6}}\put(98,16){\Huge $j$}
      \put(97.12,7.88){\line(1,-1){15.76}}
      \thinlines
      \put(65,-2){\vector(1,0){60}}
      \put(58,3){\Huge$u$\Large$+$\Huge$\lambda$}
      \put(152,-11){\huge $=\delta_{ac}{[}\theta_{2}\theta_{3}\theta_{4}{]}(0)
                   G_{a}\theta_{2}(2u)$}
      \end{picture}
      }
      \put(0,-50){\Huge {\rm Fig}.\ 7 \ Graphical representations}
      \put(45,-60){\Huge of unitary  relations between $K$'s}
\end{picture}



\setlength{\unitlength}{1mm}

\begin{picture}(100,100)

 \put(10,10){
     \setlength{\unitlength}{0.7mm}
     \begin{picture}(50,50)
     \thicklines
     \put(4,0){\line(1,0){42}}
     \put(4,50){\line(1,0){42}}
     \put(0,4){\line(0,1){42}}
     \put(2.83,2.83){\line(1,1){44.34}}
     \put(2.83,47.17){\line(1,-1){44.34}}
     \put(0,0){\circle{8}}    \put(-10,-10){\Huge $a$}
     \put(0,50){\circle{8}}   \put(-10,55){\Huge $b$}
     \put(50,0){\circle{8}}   \put(60,-10){\Huge $c$}
     \put(50,50){\circle{8}}  \put(60,55){\Huge $d$}
     \put(25,25){\circle*{8}} \put(33,24){\Huge $j$}
     \thinlines
     \put(25,-25){\vector(0,1){100}} \put(23,-31){\Huge $w$}
     \put(-14,39){\vector(1,-1){23}} \put(-21,39){\Huge $u$}
     \put(-14,11){\vector(1,1){23}}  \put(-21,8){\Huge $v$}
     \put(9,16){\vector(1,0){50}}
     \put(9,34){\vector(1,0){50}}
     \end{picture}
 }
 \put(63,25){\Huge $=$}
 \put(80,10){
     \setlength{\unitlength}{0.7mm}
     \begin{picture}(50,50)
     \thicklines
     \put(4,0){\line(1,0){42}}
     \put(4,50){\line(1,0){42}}
     \put(50,4){\line(0,1){42}}
     \put(2.83,2.83){\line(1,1){44.34}}
     \put(2.83,47.17){\line(1,-1){44.34}}
     \put(0,0){\circle{8}}    \put(-10,-10){\Huge $a$}
     \put(0,50){\circle{8}}   \put(-10,55){\Huge $b$}
     \put(50,0){\circle{8}}   \put(60,-10){\Huge $c$}
     \put(50,50){\circle{8}}  \put(60,55){\Huge $d$}
     \put(25,25){\circle*{8}} \put(13,24){\Huge $j$}
     \thinlines
     \put(25,-25){\vector(0,1){100}} \put(23,-31){\Huge $w$}
     \put(41,34){\vector(1,-1){23}} \put(64,6){\Huge $u$}
     \put(41,16){\vector(1,1){23}}  \put(64,40){\Huge $v$}
     \put(-9,16){\vector(1,0){50}}
     \put(-9,34){\vector(1,0){50}}
     \end{picture}
 }

\put(-5,-70){\Huge {\rm Fig}.\ 8 \ Graphical representation of
            (\ref{eq:defW})}

\end{picture}



\setlength{\unitlength}{1mm}

\begin{picture}(145,230)
 \put(60,160){
      \begin{picture}(40,40)
      \thicklines
      \put(3,20){\line(1,0){34}}
      \put(0,20){\circle*{8}} \put(-4,24){\Huge $b$}
      \put(2.12,22.12){\line(1,1){15.73}}
      \put(40,20){\circle{6}} \put(46,18){\Huge $c$}
      \put(2.12,17.88){\line(1,-1){15.73}}
      \put(20,0){\circle{6}} \put(24,-9){\Huge $i$}
      \put(37.88,22.12){\line(-1,1){15.73}}
      \put(20,40){\circle{6}} \put(24,46){\Huge $j$}
      \put(37.88,17.88){\line(-1,-1){15.73}}
      \thinlines
      \put(-8,8){\vector(1,0){55}}
      \put(-8,32){\vector(1,0){55}}
      \thicklines
      \put(-37.5,21){\line(1,0){36}}
      \put(-40,20){\circle{6}} \put(-49,18){\Huge $a$}
      \put(-37.5,19){\line(1,0){36}}
      \thinlines
      \put(20,-20){\vector(0,1){80}} \put(18,-26){\Huge $w$}
      \put(-30,30){\vector(1,-1){22}} \put(-33,33){\Huge $u$}
      \put(-30,10){\vector(1,1){22}} \put(-33,5){\Huge $v$}
      \end{picture}
              }
 \put(15,88){\Huge $=$}
 \put(40,70){
      \begin{picture}(40,40)
      \thicklines
      \put(3,20){\line(1,0){34}}
      \put(0,20){\circle{6}} \put(-8,18){\Huge $a$}
      \put(2.12,22.12){\line(1,1){15.73}}
      \put(40,20){\circle*{8}} \put(44,24){\Huge $b$}
      \put(2.12,17.88){\line(1,-1){15.73}}
      \put(20,0){\circle{6}}  \put(24,-9){\Huge $i$}
      \put(37.88,22.12){\line(-1,1){15.73}}
      \put(20,40){\circle{6}} \put(24,46){\Huge $j$}
      \put(37.88,17.88){\line(-1,-1){15.73}}
      \thinlines
      \put(-7,32){\vector(1,0){55}}
      \put(-7,8){\vector(1,0){55}}
      \thicklines
      \put(42.5,21){\line(1,0){35}}
      \put(80,20){\circle{6}} \put(86,18){\Huge $c$}
      \put(42.5,19){\line(1,0){35}}
      \thinlines
      \put(20,-20){\vector(0,1){80}} \put(18,-26){\Huge $w$}
      \put(48,32){\vector(1,-1){22}} \put(73,7){\Huge $u$}
      \put(48,8){\vector(1,1){22}}   \put(73,33){\Huge $v$}
      \end{picture}
              }
\put(-5,20){\Huge {\rm Fig}.\ 9
             \ Graphical representation of (\ref{eq:defWb})}

\end{picture}


\setlength{\unitlength}{1mm}

\begin{picture}(145,230)
      \multiput(90,50)(0,40){2}
         {
         \begin{picture}(40,40)
           \thicklines
           \put(3,20){\line(1,0){34}}
           \put(0,20){\circle{6}}
           \put(2.12,17.88){\line(1,-1){15.76}}
           \put(20,0){\circle*{6}}
           \put(22.12,2.12){\line(1,1){15.76}}
           \put(40,20){\circle{6}}
           \put(2.12,22.12){\line(1,1){15.76}}
           \put(20,40){\circle*{6}}
           \put(22.12,37.88){\line(1,-1){15.76}}
           \thinlines
           \put(-20,10){\vector(1,0){80}}
           \put(-26,8){\Huge $u$}
           \put(-20,30){\vector(1,0){80}}
           \put(-26,30){\Huge $v$}
         \end{picture}
          }
      \put(90,170)
         {
         \begin{picture}(40,40)
           \thicklines
           \put(3,20){\line(1,0){34}}
           \put(0,20){\circle{6}}
           \put(2.12,17.88){\line(1,-1){15.76}}
           \put(20,0){\circle*{6}}
           \put(22.12,2.12){\line(1,1){15.76}}
           \put(40,20){\circle{6}}
           \put(2.12,22.12){\line(1,1){15.76}}
           \put(20,40){\circle*{6}}
           \put(22.12,37.88){\line(1,-1){15.76}}
           \thinlines
           \put(-20,10){\vector(1,0){80}}
           \put(-26,8){\Huge $u$}
           \put(-20,30){\vector(1,0){80}}
           \put(-26,30){\Huge $v$}
         \end{picture}
          }
      \thinlines
      \put(111,30){\vector(0,1){200}}
      \put(108,24){\Huge $w$}
      \put(73,70){\Huge $a$\Large$\mbox{}_{M-1}$}
      \put(73,110){\Huge $a$\Large$\mbox{}_{M-2}$}
      \put(78,190){\Huge $a_{0}$}
      \put(137,69){\Huge $b$\Large$\mbox{}_{M-1}$}
      \put(137,109){\Huge $b$\Large$\mbox{}_{M-2}$}
      \put(137,189){\Huge $b_{0}$}
      \put(117,48){\Huge $i_{0}$}
      \put(117,88){\Huge $i$\Large$\mbox{}_{M-1}$}
      \put(117,128){\Huge $i$\Large$\mbox{}_{M-2}$}
      \put(117,168){\Huge $i_{1}$}
      \put(117,208){\Huge $i_{0}$}
      \multiput(80,117)(0,4){17}{\circle*{0.8}}
      \multiput(140,117)(0,4){17}{\circle*{0.8}}
      \multiput(120,137)(0,4){8}{\circle*{0.8}}
      \put(-15,130){\Huge ${\cal L}(u,v,w)_{a_{0}a_{1}\cdots a_{M-1}}
                    ^{b_{0}b_{1}\cdots b_{M-1}} =$}
      \put(-15,0){\Huge {\rm Fig}.\ 10 \ Graphical representations
                 of ${\cal L}(u,v,w)$}
\end{picture}



\setlength{\unitlength}{1mm}
\begin{picture}(145,230)
    \put(20,170)
    {
       \multiput(0,0)(0,20){2}
         {\setlength{\unitlength}{0.5mm}
         \begin{picture}(40,40)
           \thicklines
           \put(3,20){\line(1,0){34}}
           \put(0,20){\circle{6}}
           \put(2.12,17.88){\line(1,-1){15.76}}
           \put(20,0){\circle*{6}}
           \put(22.12,2.12){\line(1,1){15.76}}
           \put(40,20){\circle{6}}
           \put(2.12,22.12){\line(1,1){15.76}}
           \put(20,40){\circle*{6}}
           \put(22.12,37.88){\line(1,-1){15.76}}
           \thinlines
           \put(-20,10){\vector(1,0){100}}
           \put(-29,8){\Large $u$}
           \put(-20,30){\vector(1,0){100}}
           \put(-29,30){\Large $v$}
           \put(20,-20){\vector(0,1){80}}
         \end{picture}
         }
    \put(0,0)
          {\setlength{\unitlength}{0.5mm}
         \begin{picture}(40,40)
         \put(-15,18){\Large $a_{1}$}
         \put(-15,58){\Large $a_{0}$}
         \end{picture}
          }
       \multiput(20,0)(0,20){2}
         {\setlength{\unitlength}{0.5mm}
         \begin{picture}(40,40)
           \thicklines
           \put(0,20){\circle*{6}}
           \put(2.12,17.88){\line(1,-1){15.76}}
           \put(20,0){\circle{6}}
           \put(1.08,23.0){\line(1,1){15.95}}
           \put(3.0,21.12){\line(1,1){15.86}}
           \put(20,40){\circle{6}}
           \thinlines
           \put(10,-20){\vector(0,1){80}}
         \end{picture}
          }
       \put(10,-15){\Large $w'$}
       \put(25,-15){\Large $w$}
    \put(20,0)
          {\setlength{\unitlength}{0.5mm}
         \begin{picture}(40,40)
         \put(25,-2){\Large $c_{0}$}
         \put(25,38){\Large $c_{1}$}
         \put(25,78){\Large $c_{0}$}
         \put(-35,-2){\Large $i_{0}$}
         \put(-35,38){\Large $i_{1}$}
         \put(-35,78){\Large $i_{0}$}
         \end{picture}
          }
     }
\put(68,180){\Huge $=$}

    \put(90,170)
    {
       \multiput(0,-20)(0,20){3}
         {\setlength{\unitlength}{0.5mm}
         \begin{picture}(40,40)
           \thicklines
           \put(3,20){\line(1,0){34}}
           \put(0,20){\circle{6}}
           \put(2.12,17.88){\line(1,-1){15.76}}
           \put(20,0){\circle*{6}}
           \put(22.12,2.12){\line(1,1){15.76}}
           \put(40,20){\circle{6}}
           \put(2.12,22.12){\line(1,1){15.76}}
           \put(20,40){\circle*{6}}
           \put(22.12,37.88){\line(1,-1){15.76}}
           \thinlines
         \end{picture}
         }
       \put(12,-30){\vector(0,1){80}}
       \put(10,53){\Large $w'$}
       \put(-10,-5){\vector(1,0){30}}
       \put(-15,-5){\Large $w$}
       \put(20,-2.5){\oval(12.6,5)[rb]}
       \put(26.5,-3.0){\vector(0,1){50}}
       \put(-10,-15){\vector(1,0){60}}
       \put(-20,-22){\Large $w+\lambda$}
    \put(0,0)
          {\setlength{\unitlength}{0.5mm}
         \begin{picture}(40,40)
         \put(-15,-25){\Large $a'$}
         \put(-15,18){\Large $a_{1}$}
         \put(-15,58){\Large $a_{0}$}
         \end{picture}
          }
       \multiput(20,0)(0,20){2}
         {\setlength{\unitlength}{0.5mm}
         \begin{picture}(40,40)
           \thicklines
           \put(0,20){\circle*{6}}
           \put(2.12,17.88){\line(1,-1){15.76}}
           \put(20,0){\circle{6}}
           \put(1.08,23.0){\line(1,1){15.95}}
           \put(3.0,21.12){\line(1,1){15.86}}
           \put(20,40){\circle{6}}
           \thinlines
           \put(-60,10){\vector(1,0){100}}
           \put(46,8){\Large $u$}
           \put(-60,30){\vector(1,0){100}}
           \put(46,30){\Large $v$}
         \end{picture}
          }
    \put(20,0)
          {\setlength{\unitlength}{0.5mm}
         \begin{picture}(40,40)
         \put(25,-2){\Large $c_{0}$}
         \put(25,38){\Large $c_{1}$}
         \put(25,78){\Large $c_{0}$}
         \put(5,-22){\Large $c_{0}$}
         \put(-35,-45){\Large $i'$}
         \put(-35,-2){\Large $i_{0}$}
         \put(-35,38){\Large $i_{1}$}
         \put(-35,78){\Large $i'$}
         \put(-75,-80){{\Large $\times$}$\Biggl(\dfrac{G_{a'}}
                      {{[}\theta_{2}\theta_{3}\theta_{4}{]}(0)
                       G_{c_{0}}\theta_{2}(2w-2w')}\Biggr)$}
         \end{picture}
          }
      }

\put(138,180){\Huge $=$}
    \put(-10,55)
    {
       \multiput(0,0)(0,20){2}
         {\setlength{\unitlength}{0.5mm}
         \begin{picture}(40,40)
           \thicklines
           \put(0,20){\circle{6}}
           \put(2.12,17.88){\line(1,-1){15.76}}
           \put(20,0){\circle*{6}}
           \put(1.08,23.0){\line(1,1){15.95}}
           \put(3.0,21.12){\line(1,1){15.86}}
           \put(20,40){\circle*{6}}
           \thinlines
           \put(-20,10){\vector(1,0){80}}
           \put(-26,8){\Large $u$}
           \put(-20,30){\vector(1,0){80}}
           \put(-26,30){\Large $v$}
         \end{picture}
          }
         \put(6,-5){\vector(0,1){45}}
         \put(10,40){\oval(8,11)[lt]}
         \put(10,45.5){\vector(1,0){25}}
         \put(38,45.5){\Large $w$}
    \put(0,0)
          {\setlength{\unitlength}{0.5mm}
         \begin{picture}(40,40)
         \put(-15,18){\Large $a_{1}$}
         \put(-15,58){\Large $a_{0}$}
         \end{picture}
          }
       \multiput(10,-10)(0,20){3}
         {\setlength{\unitlength}{0.5mm}
         \begin{picture}(40,40)
           \thicklines
           \put(3,20){\line(1,0){34}}
           \put(0,20){\circle{6}}
           \put(2.12,17.88){\line(1,-1){15.76}}
           \put(20,0){\circle*{6}}
           \put(22.12,2.12){\line(1,1){15.76}}
           \put(40,20){\circle{6}}
           \put(2.12,22.12){\line(1,1){15.76}}
           \put(20,40){\circle*{6}}
           \put(22.12,37.88){\line(1,-1){15.76}}
           \thinlines
         \end{picture}
         }
       \put(7,-28){\setlength{\unitlength}{0.5mm}
         \begin{picture}(40,40)
         \put(10,45.5){\vector(1,0){40}}
         \put(48.5,36.5){\Large $w+\lambda$}
         \put(25,25){\vector(0,1){145}}
         \put(22,172){\Large $w'$}
         \end{picture}
        }
    \put(20,0)
          {\setlength{\unitlength}{0.5mm}
         \begin{picture}(40,40)
         \put(25,-2){\Large $c_{0}$}
         \put(25,38){\Large $c_{1}$}
         \put(25,78){\Large $c_{0}$}
         \put(-28,-15){\Large $a'$}
         \put(-23,93){\Large $b_{0}$}
         \put(8,-25){\Large $i'$}
         \put(8,18){\Large $i_{1}$}
         \put(8,58){\Large $i_{0}$}
         \put(8,98){\Large $i'$}
         \put(-75,-52){{\Large $\times$}$\Biggl(\dfrac{G_{a'}}
                      {{[}\theta_{2}\theta_{3}\theta_{4}{]}(0)
                       G_{c_{0}}\theta_{2}(2w-2w')}\Biggr)$}
         \end{picture}
          }
     }
\put(30,73){\Huge $=$}
\put(50,55)
   {
    \multiput(0,0)(0,20){2}
         {\setlength{\unitlength}{0.5mm}
         \begin{picture}(40,40)
           \thicklines
           \put(0,20){\circle{6}}
           \put(2.12,17.88){\line(1,-1){15.76}}
           \put(20,0){\circle*{6}}
           \put(1.08,23.0){\line(1,1){15.95}}
           \put(3.0,21.12){\line(1,1){15.86}}
           \put(20,40){\circle*{6}}
           \thinlines
           \put(-18,10){\vector(1,0){90}}
           \put(-26,8){\Large $u$}
           \put(-18,30){\vector(1,0){90}}
           \put(-26,30){\Large $v$}
         \end{picture}
          }
           \put(5,0){\vector(0,1){45}}
           \put(2,48){\Large $w$}
    \put(0,0)
          {\setlength{\unitlength}{0.5mm}
         \begin{picture}(40,40)
         \put(-15,18){\Large $a_{1}$}
         \put(-15,58){\Large $a_{0}$}
         \end{picture}
          }

       \multiput(10,-20)(0,20){3}
         {\setlength{\unitlength}{0.5mm}
         \begin{picture}(40,40)
           \thicklines
           \put(3,0){\line(1,0){34}}
           \put(0,0){\circle*{6}}
           \put(3,40){\line(1,0){34}}
           \put(0,40){\circle{6}}
           \put(2.12,2.12){\line(1,1){35.76}}
           \put(2.12,37.88){\line(1,-1){35.76}}
           \put(20,20){\circle*{6}}
           \put(40,0){\circle{6}}
           \put(40,40){\circle{6}}
         \end{picture}
          }
       \put(7,-33){\setlength{\unitlength}{0.5mm}
         \begin{picture}(40,40)
         \put(10,55){\vector(1,0){40}}
         \put(52,50.5){\Large $w+\lambda$}
         \put(10,35){\vector(1,0){40}}
         \put(52,33.0){\Large $w$}
         \put(25,15){\vector(0,1){140}}
         \put(22,159){\Large $w'$}
         \end{picture}
        }
    \put(20,0)
          {\setlength{\unitlength}{0.5mm}
         \begin{picture}(40,40)
         \put(25,-45){\Large $c_{0}$}
         \put(25,-2){\Large $c_{0}$}
         \put(25,38){\Large $c_{1}$}
         \put(25,78){\Large $c_{0}$}
         \put(8,-25){\Large $i'$}
         \put(8,18){\Large $i_{1}$}
         \put(8,58){\Large $i_{0}$}
         \put(-32,-13){\Large $a'$}
         \put(-33,-43){\Large $b_{0}$}
         \put(-23,88){\Large $b_{0}$}
         \put(-63,-67){{\Large $\times$}$\Biggl(\dfrac{G_{a'}}
                      {{[}\theta_{2}\theta_{3}\theta_{4}{]}(0)
                       G_{c_{0}}\theta_{2}(2w-2w')}\Biggr)$}
         \end{picture}
          }
     }

\put(93,73){\Huge $=$}
\put(110,55)
   {
    \multiput(0,0)(0,20){2}
         {\setlength{\unitlength}{0.5mm}
         \begin{picture}(40,40)
           \thicklines
           \put(0,20){\circle{6}}
           \put(2.12,17.88){\line(1,-1){15.76}}
           \put(20,0){\circle*{6}}
           \put(1.08,23.0){\line(1,1){15.95}}
           \put(3.0,21.12){\line(1,1){15.86}}
           \put(20,40){\circle*{6}}
           \thinlines
           \put(-20,10){\vector(1,0){100}}
           \put(86,8){\Large $u$}
           \put(-20,30){\vector(1,0){100}}
           \put(86,30){\Large $v$}
         \end{picture}
          }
           \put(5,0){\vector(0,1){45}}
           \put(2,48){\Large $w$}
    \put(0,0)
          {\setlength{\unitlength}{0.5mm}
         \begin{picture}(40,40)
         \put(-15,18){\Large $a_{1}$}
         \put(-15,58){\Large $a_{0}$}
         \end{picture}
          }
       \multiput(10,0)(0,20){2}
         {\setlength{\unitlength}{0.5mm}
         \begin{picture}(40,40)
           \thicklines
           \put(3,0){\line(1,0){34}}
           \put(0,0){\circle{6}}
           \put(3,40){\line(1,0){34}}
           \put(0,40){\circle{6}}
           \put(2.12,2.12){\line(1,1){35.76}}
           \put(2.12,37.88){\line(1,-1){35.76}}
           \put(20,20){\circle*{6}}
           \put(40,0){\circle{6}}
           \put(40,40){\circle{6}}
         \end{picture}
          }
       \put(7,-33){\setlength{\unitlength}{0.5mm}
         \begin{picture}(40,40)
         \put(25,45){\vector(0,1){110}}
         \put(22,159){\Large $w'$}
         \end{picture}
        }
    \put(20,0)
          {\setlength{\unitlength}{0.5mm}
         \begin{picture}(40,40)
         \put(25,-2){\Large $c_{0}$}
         \put(25,38){\Large $c_{1}$}
         \put(25,78){\Large $c_{0}$}
         \put(5,18){\Large $i_{1}$}
         \put(5,58){\Large $i_{0}$}
         \put(-20,-13){\Large $b_{0}$}
         \put(-20,88){\Large $b_{0}$}
         \end{picture}
        }
   \put(-125,-55){\huge {\rm Fig}.\ 11
                  \ Graphical representation of a proof}
   \put(-120,-65) {\huge of ${\cal L}(u,v,w')$ $\Phi(u,v,w)$
                          $=$ $\Phi(u,v,w)$ ${\cal L}(v,u,w')$}
   }
\end{picture}


\begin{thebibliography}{99}

\bibitem{FZ82}
V.A. Fateev and A.B. Zamolodchikov,
Self-dual solutions of star-triangle relations
       in {$Z\mbox{}_{N}$}-model,
Phys. Lett. A92,
pp. 37-39,
1982.

\bibitem{AMcPTY87}
H. Au-Yang{,} B.M. McCoy{,} J.H.H. Perk{,} S. Tang and M. Yan,
Commuting transfer matrices in the chiral {P}otts model:
       solutions of star-triangle equations with genus {$>$} 1,
Phys. Lett. A123,
pp219-223,
1987.

\bibitem{BPAuY88}
R.J. Baxter{,} J.H.H. Perk and H. Au-Yang,
New solution of the star-triangle relations for the
       chiral {P}otts model,
Phys. Lett. A128,
pp. 138-142,
1988.

\bibitem{AuYP89}
H. Au-Yang and J.H.H. Perk,
Onsager's star-triangle equation: Master key to
       integrability,
Adv. Stud. in Pure Math. 19,
pp.57-94,
1989.

\bibitem{BBP89}
R.J. Baxter{,} V.V. Bazhanov and J.H.H. Perk,
Functional relations for transfer matrices of the chiral
        {P}otts model,
Int. J. Mod. Phys. B4,
pp.803,
1989.

\bibitem{Baxter88-2}
R.J. Baxter,
Free energy of solvable chiral {P}otts model,
J. Stat. Phys. 52,
pp.639,
1988.

\bibitem{AMcP89-3}
G. Albertini{,} B.M. McCoy and J.H.H. Perk,
Eigenvalue spectrum of superintegrable chiral {P}otts model,
Adv. Stud. in Pure Math. 19,
pp.1-55,
1989.

\bibitem{Baxter90},
R.J. Baxter,
Chiral {P}otts model, eigenvalues of transfer matrix,
Phys. Lett. A146,
110-114,
1990.

\bibitem{ADMc91}
G. Albertini{,} S. Dasmahapatra and B.M. McCoy,
Spectrum and Completeness of the integrable
       3-state {P}otts model: a finite size study,
Int. J. Mod. Phys. A7. Suppl. 1A,
pp1-53,
1992.

\bibitem{Albertini92}
G. Albertini,
Bethe-{A}nsatz type equation for the {F}ateev-{Z}amolodchikov
       spin model,
J. Phys: Math. Gen. 25,
pp. 1799-1813,
1992.

\bibitem{KM86}
M. Kashiwara and T. Miwa,
A class of elliptic solutions to the star-triangle relations,
Nucl. Phys. B275 [FS17],
pp. 121-134,
1986.

\bibitem{HY90}
K. Hasegawa and Y. Yamada,
Algebraic derivation of broken {$Z\mbox{}_{N}$}
       symmetric model,
Phys. Lett. A 146,
pp. 387-396,
1990.

\bibitem{Sklyanin82-1}
E. Sklyanin,
Some algebraic structure connected with the {Y}ang-{B}axter equation,
Funk. Anal. Appl. 16 No4,
pp. 27-34,
1982.

\bibitem{Sklyanin82-2}
E. Sklyanin,
Some algebraic structure connected with the {Y}ang-{B}axter equation,
       representation of quantum algebra,
Funk. Anal. Appl. 16 No4,
pp. 34-48,
1982.

\bibitem{Baxter71}
R.J. Baxter,
Eight-vertex model in lattice statistics,
Phys. Rev. Lett. 26,
pp.832-833,
1971.

\bibitem{Baxter72}
R.J. Baxter,
Partition function of eight-vertex model,
Ann. Phys. (N.Y.) 70,
pp.193-228,
1972.

\bibitem{Bible82}
R.J. Baxter,
 ``Exactly {S}olved {M}odels in {S}tatistical {M}echanics'',
Academic Press, London,
1982.

\bibitem{BS90}
V. V. Bazhanov and Yu. G. Stroganov,
Chiral {P}otts model as a descendent of the six-vertex model,
J. Stat. Phys., Vol. 59, Nos. 3/4,
pp.799-817,
1990.

\bibitem{BR89}
V. V. Bazhanov and N. Yu. Reshetikhin,
Critical {RSOS} models and conformal field theory,
Int. J. Mod. Phys. A, Vol 4, No. 1,
pp.115-142,
1989.

\bibitem{JMO86}
M. Jimbo{,} T. Miwa and M. Okado,
Solvable lattice model with broken {$Z\mbox{}_{N}$}
       symmetry and {H}ecke's indefinite modular form,
Nucl. Phys. B275 [FS17],
pp. 517-545,
1986.

\bibitem{Baxter73-1}
R.J. Baxter,
Eight-vertex model in lattice statistics and
      one-dimensional anisotropic {H}eisenberg chain {I},
      some fundamental eigenvectors,
Ann. Phys. (N.Y.) 76,
pp.1-24,
1973.

\bibitem{Baxter73-2}
R.J. Baxter,
Eight-vertex model in lattice statistics and
      one-dimensional anisotropic {H}eisenberg chain {II},
      equivalence to a generalized ice-type lattice model,
Ann. Phys. (N.Y.) 76,
pp.25-47,
1973.

\bibitem{Baxter73-3}
R.J. Baxter,
Eight-vertex model in lattice statistics and
      one-dimensional anisotropic {H}eisenberg chain {III},
      eigenvalues of transfer matrix and Hamiltonian,
Ann. Phys. (N.Y.) 76,
pp.48-71,
1973.

\bibitem{Pasquier88}
V.Pasquier,
Etiology of {IRF} model,
Comm. Math. Phys. 118,
pp.357-364,
1988.

\bibitem{Hasegawa93}
K. Hasegawa,
Crossing symmetry in elliptic solutions of the
       {Y}ang-{B}axter equation and a new {L}-operator for
       {B}elavin's solution,
J. Phys. A26,
pp.3211,
1993.

\bibitem{Quano92}
Y.-H. Quano,
``Sklyanin algebra and solvable lattice model''
in the proceedings of the {W}orkshop on
         {``}{\it {N}ew {A}spects of {Q}uantum {F}ield {T}heories}{``}
         {,} {S}. {N}ojiri ed.{,} {INT-T-513},
pp. 29-44,
1992.

\bibitem{Hasegawa92}
K. Hasegawa,
On the crossing symmetry of broken {${\bf Z}_{N}$}-symmetric
        solutions of the {Y}ang-{B}axter equation,
in {``}{\it {R}epresentation theory of {L}ie groups
         and {L}ie algebras}{``}{,}
         the proceedings of the {K}awaguchi-ko conference{,}
         {T}.{K}awazoe{,} {T},{O}shima and {S}.{S}ano eds{,}
         {W}orld {S}cientific{,} {S}ingapore,
pp.22-58,
1992.

\bibitem{TS72}
M. Takahashi and M. Suzuki,
One-dimensional anisotropic {H}eisenberg model
      at finite temperature,
Prog. Theo. Phys. vol48, 6B,
pp.2187-2209,
1972.

\bibitem{Gantmacher59}
F.R. Gantmacher,
''Matrix Theory'',
Chelsea Publishing Co. New York,
1959.

\bibitem{Mumford83}
D. Mumford,
``Tata Lectures on Theta {I}'',
Birkhauser, Boston,
1983.

\bibitem{WW58}
E. T. Whittaker and G.N. Watson,
``A {C}ourse in {M}odern {A}nalysis'', 4th ed.,
Cambridge Univ. Press,
1958.












\end{thebibliography}
\end{document}